\documentclass[useAMS,usenatbib]{mn2e}

\usepackage{graphicx}
\usepackage{amsmath}
\usepackage{mathrsfs}
\usepackage[]{txfonts}
\usepackage{subfigure}
\usepackage{epsfig}
\usepackage{booktabs}
\usepackage{url}
\usepackage{xcolor}

\usepackage[colorinlistoftodos]{todonotes}

\def\simless{\mathbin{\lower 3pt\hbox
{$\rlap{\raise 5pt\hbox{$\char'074$}}\mathchar"7218$}}}   
\def\simmore{\mathbin{\lower 3pt\hbox
{$\rlap{\raise 5pt\hbox{$\char'076$}}\mathchar"7218$}}}   
\newcommand{\be}{\begin{equation}}
\newcommand{\ee}{\end{equation}}
\newcommand       \bea          {\begin{eqnarray}}
\newcommand       \eea          {\end{eqnarray}}

\def\simlt{\mathrel{\hbox{\rlap{\hbox{\lower4pt\hbox{$\sim$}}}\hbox{$<$}}}}
\def\simgt{\mathrel{\hbox{\rlap{\hbox{\lower4pt\hbox{$\sim$}}}\hbox{$>$}}}}
\def\lesssim{\mathrel{\hbox{\rlap{\hbox{\lower4pt\hbox{$\sim$}}}\hbox{$<$}}}}

\def\gtrsim{\mathrel{\hbox{\rlap{\hbox{\lower4pt\hbox{$\sim$}}}\hbox{$>$}}}}

\title[$r$-process in NS merger remnant disc outflows]{Production of
  the entire range of $r$-process nuclides by black hole accretion disc outflows from
  neutron star mergers} 

\author[Wu, Fern\'andez, Mart\'inez-Pinedo, \& Metzger]{Meng-Ru
  Wu$^{1}$$\thanks{E-mail:mwu@theorie.ikp.physik.tu-darmstadt.de}$,
  Rodrigo Fern\'andez$^{2,3}$, Gabriel Mart{\'{\i}}nez-Pinedo$^{1,4}$, 
\vspace{0.25cm}
\\{\LARGE\rm Brian~D.~Metzger$^{5}$}\\
$^{1}$ Institut f\"{u}r Kernphysik, Technische Universit\"{a}t Darmstadt, 64289 Darmstadt, Germany \\
$^{2}$ Department of Physics, University of California, Berkeley, CA 94720, USA\\
$^{3}$ Department of Astronomy and Theoretical Astrophysics Center, University of California, Berkeley, CA 94720, USA\\
$^{4}$ GSI Helmholtzzentrum f\"{u}r Schwerionenforschung, Planckstr.~1, 64291 Darmstadt, Germany\\
$^{5}$Department of Physics and Columbia Astrophysics Laboratory, Columbia University, New York, NY, 10027, USA\\
}

\begin{document}
\date{Received / Accepted}
\pagerange{\pageref{firstpage}--\pageref{lastpage}} \pubyear{2016}

\maketitle
\label{firstpage}

\begin{abstract}
  We consider $r$-process nucleosynthesis in outflows from black hole
  accretion discs formed in double neutron star and neutron star --
  black hole mergers.  These outflows, powered by angular momentum
  transport processes and nuclear recombination, represent an
  important -- and in some cases dominant -- contribution to the total
  mass ejected by the merger.  Here we calculate the nucleosynthesis
  yields from disc outflows using thermodynamic trajectories from
  hydrodynamic simulations, coupled to a nuclear reaction network.  We
  find that outflows produce a robust abundance pattern around the
  second $r$-process peak (mass number $A \sim 130$), independent of model
  parameters, with significant production of $A < 130$ nuclei. This
  implies that dynamical ejecta with high electron fraction may not be
  required to explain the observed abundances of $r$-process elements
  in metal poor stars.  Disc outflows reach the third peak
  ($ A \sim 195$) in most of our simulations, although the amounts
  produced depend sensitively on the disc viscosity, initial mass or
  entropy of the torus, and nuclear physics inputs.  Some of our
  models produce an abundance spike at $A = 132$ that is absent in the
  Solar system $r$-process distribution. The spike arises from
  convection in the disc and depends on the treatment of nuclear
  heating in the simulations. We conclude that disc outflows provide
  an important -- and perhaps dominant -- contribution to the
  $r$-process yields of compact binary mergers, and hence must be
  included when assessing the contribution of these systems to the
  inventory of $r$-process elements in the Galaxy.
\end{abstract} 
  

\section{Introduction}
\label{sec-intro}

Approximately half of the elements with mass number $A > 70$, and all
of the transuranic nuclei, are formed by the rapid neutron capture
process (the $r$-process; \citealt{Burbidge+57}, \citealt{Cameron57}).
The astrophysical site of this process has been under debate for more
than 50 yrs (see, e.g., \citealt{Qian&Wasserburg07}, \citealt{Arnould+07}, 
\citealt{Sneden+08}, \citealt{Thielemann+11} for
reviews).  Neutrino-driven outflows from proto neutron stars (NSs)
following core collapse supernovae have for long been considered the
prime candidate site (\citealt{Meyer+92}; \citealt{Woosley+94};
\citealt{Qian&Woosley96}). However, state-of-the-art calculations find
thermodynamic conditions that are
at best marginal for the $r$-process, especially when extending up to the
heaviest third-peak elements with mass number $A \sim 195$
(e.g.~\citealt{MartinezPinedo+12}; \citealt{Roberts+12};
\citealt{martinez2014}).  

Prospects for a successful $r$-process in neutrino-driven
outflows may be improved if the proto-NS is born with a strong
magnetic field and very rapid rotation
(e.g.~\citealt{Thompson+04,Metzger+07,Vlasov+14}).  If the supernova
itself is MHD-powered, additional magnetocentrifugal acceleration
could substantially reduce the electron fraction of the outflow
compared to its value in the purely neutrino-driven case favouring
the occurrence of an $r$-process (e.g.~\citealt{Burrows+07}, \citealt{Winteler.Kaeppeli.ea:2012},
\citealt{Nishimura+15}). However, current simulations of
MHD-powered supernova explosions need further improvements, especially
considering the role of instabilities on the jet structure which
manifest in three dimensions (\citealt{Moesta+14}).

The coalescence of double NS (NS--NS) and NS-black hole (NS--BH)
binaries \citep{Lattimer&Schramm74} provides an alternative
$r$-process source.  Numerical simulations of these events show that a
robust outcome of the merger is the ejection of
$\sim 10^{-4}$--$10^{-1}$~$M_\odot$ of highly neutron-rich matter on
the dynamical time (e.g.~\citealt{Hotokezaka+13},
\citealt{Bauswein+13}; see \citealt{lehner2014} and
\citealt{Baiotti:2016qnr} for recent reviews).  Estimates show that
NS--NS/NS--BH mergers could contribute a sizable fraction of the total
production of $r$-process elements in the Galaxy, depending on the
uncertain merger rates.  At the same time, previous arguments against
mergers being dominant $r$-process sites based on Galactic chemical
evolution and the observed prompt enrichment of $r$-process nuclei in
metal poor stars (e.g.~\citealt{Argast+04}) have been challenged
(e.g., \citealt{vdvoort2015,shen2015,hirai2015}).  Additional evidence
supporting the presence of a `high yield' $r$-process site -- like an
NS--NS/NS--BH merger -- includes the discovery of highly $r$-process
enriched stars in the ultra-faint dwarf galaxy Reticulum II
(\citealt{Ji+16}), and the abundance of the short-lived isotope
$^{244}$Pu on the sea floor (\citealt{Wallner+15, Hotokezaka+15}).

Nucleosynthesis in NS--NS/NS--BH mergers has also received a recent surge 
of interest due to the realization that the radioactive decay of the 
$r$-process ejecta can power a thermal transient 
(a ``kilonova''; e.g.,~\citealt{Li&Paczynski98}, \citealt{Metzger+10},
\citealt{Roberts+11}, \citealt{Barnes&Kasen13}, \citealt{tanaka2013}), which
could serve as a promising electromagnetic counterpart to the gravitational
waves (\citealt{Metzger&Berger12}).  The detection of a possible
kilonova following the {\it Swift} GRB 130603B (\citealt{Berger+13};
\citealt{Tanvir+13}) highlights the potential of kilonovae as both a unique
diagnostic of physical processes at work during the merger
and a direct probe of the formation of $r$-process nuclei
(see, e.g., \citealt{rosswog2015}, \citealt{FM16}, \citealt{tanaka2016} 
for recent reviews).  

Previous work on the $r$-process in NS--NS/NS--BH mergers has been
focused primarily on the dynamical ejecta that is unbound promptly
during the immediate aftermath of the merger (e.g.~\citealt{Meyer89},
\citealt{Freiburghaus+99}, \citealt{goriely2005}).  Earlier
simulations that did not include weak interactions have shown this
unbound matter to be highly neutron-rich, with an electron fraction
$Y_e \lesssim 0.1$, sufficiently low to produce a robust abundance
pattern for heavy nuclei with $A \gtrsim 130$ as the result of fission
cycling (e.g., \citealt{Goriely+11}, \citealt{Korobkin+12},
\citealt{Bauswein+13}, \citealt{Mendoza-Temis.Wu.ea:2015}).  More
recently, a number of merger calculations that include the effects of
$e^\pm$ captures and neutrino irradiation in full general-relativity have 
shown that the dynamical ejecta can have a wider electron fraction
distribution($Y_e \sim 0.1-0.4$) than models without weak interaction
effects~\citep{sekiguchi2015,foucart2015a,radice2016}.  
As a result, lighter $r$-process elements with $90 \lesssim A \lesssim 130$
are generated in addition to third-peak elements \citep{wanajo2014}.
It is important to keep in mind, however, that the light element
yields in these calculations are dependent on the assumed dense-matter
equation of state and on the details of the neutrino transport
employed, in addition to the NS radii and the binary mass ratio.

\begin{figure}
\includegraphics*[width=\columnwidth]{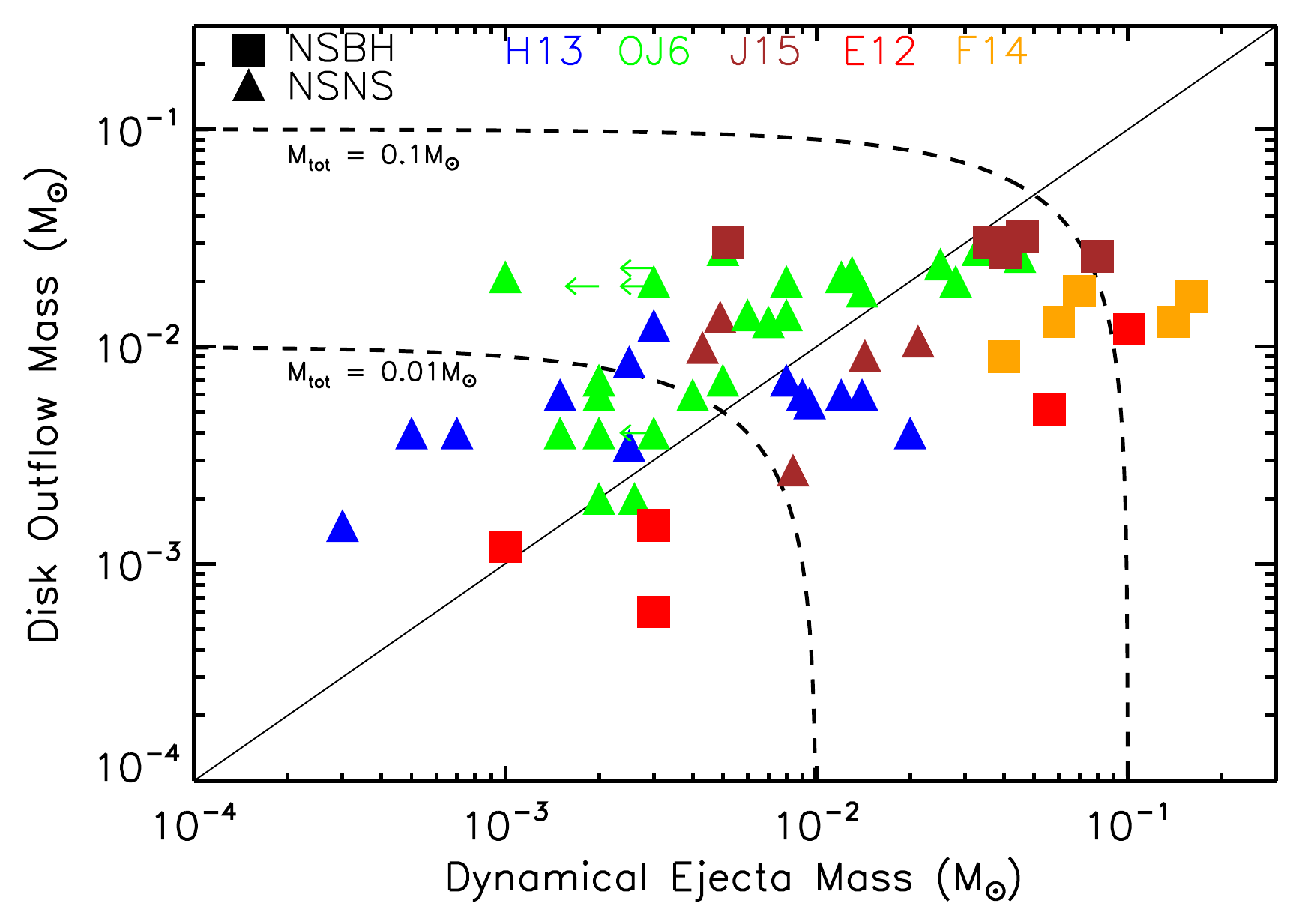}
\caption{Mass ejected dynamically during a compact binary merger
  versus that ejected in disc outflows.  Each point corresponds to the
  result of a single time-dependent NS--NS (triangles) and BH-NS
  (squares) simulation. Shown are models by \citet{Hotokezaka+13}
  (blue), \citet{Oechslin&Janka06} (green, upper limits
  shown by arrows), \citet{just2014} (brown), \citet{East+12}
  (red), and \citet{foucart2014} (orange).  The mass
  unbound in disc outflows is estimated to be 10 per cent of the mass
  of the remnant disc, based on calculations of the subsequent
  accretion disc evolution (e.g., FM13). Dashed lines show total
  ejecta mass contours (dynamical + disc winds) of $0.01 M_{\odot}$
  and $0.1M_{\odot}$, bracketing the range necessary to explain the
  Galactic production rate of heavy $r$-process nuclei
  $\sim 5\times 10^{-7}M_{\odot}$~yr$^{-1}$ \citep{Qian00}, given the
  allowed range of the rates of NS--NS mergers $\in [4,61]$~Myr$^{-1}$
  (99$\%$ confidence) calculated based on the population of Galactic
  binaries \citep{kim2015}. 
  In reality, the ejecta mass range required to reproduce
  the Galactic abundances is uncertain by greater than an order of
  magnitude, due to systematic uncertainties on the merger rate and
  depending on the precise atomic mass range under consideration 
  (e.g.,~\citealt{Bauswein:2014vfa}).
  See also \citet{FM16}.}
\label{f:brian_ejecta_mass}
\end{figure}

In addition to ejecting material dynamically, NS--NS/NS--BH mergers
result in the formation of an accretion disc around the central
remnant (e.g., \citealt{Oechslin&Janka06}); with the latter being a
promptly formed BH or a longer-lived hypermassive NS (HMNS) (e.g.,
\citealt{shibata2000}).  In both cases, the accretion disc can
generate outflows on time-scales much longer than the orbital time
(e.g., \citealt{Metzger+08, Lee+09,Metzger+09a}), and with a
contribution to the total mass ejection that can be comparable to, or even
larger than that from the dynamical ejecta
(Fig/~\ref{f:brian_ejecta_mass}, see also \citealt{FM16}).  A
relatively massive disc ($\sim 0.1M_{\odot}$) can be formed
following a NS--NS merger, as part of the process by which the
HMNS sheds angular momentum outwards prior to collapsing
into a BH (e.g.~\citealt{shibata2006}).  Long-term hydrodynamic
simulations of the disc evolution show that a significant fraction of
the initial disc mass ($\sim 5-20\%$, corresponding to
$\sim 0.01M_{\odot}$) is unbound in outflows powered by heating from
angular momentum transport and nuclear recombination, on a timescale
of $\gtrsim 1$~s (\citealt{Fernandez&Metzger13}, hereafter FM13;
\citealt{just2014}, \citealt{FKMQ14}).  As the result of weak
interactions, the electron fraction of the disc outflows
lies in the range $Y_e \sim 0.2-0.4$, generally higher than that of
the dynamical ejecta, but still sufficiently low to achieve the
$r$-process \citep{just2014}.

Although most of the previous work on merger disc wind nucleosynthesis
has focused on parametrized outflows powered by neutrino heating
(e.g.~\citealt{McLaughlin&Surman05,Surman+08,Caballero+12,Surman+14}),
in analogy with proto-NS winds, time-dependent models of the long-term
disc evolution show that neutrino heating is sub-dominant relative to
viscous heating in driving most of the disc outflow when a BH sits at
the centre (FM13, \citealt{just2014}). Neutrino heating is much more
important if the merger produces a long-lived HMNS
\citep{Dessart+09,Metzger&Fernandez14,perego2014,martin2015},
resulting in a larger ejecta mass with higher electron fraction,
depending on the uncertain lifetime of such a remnant. To date, the
only fully time-dependent nucleosynthesis study of long-term outflows
from BH accretion discs was carried out by \citet{just2014}, who
found that disc outflows can generate elements from $A\sim 80$ to the
actinides, with the contribution above $A =130$ being sensitive to
system parameters.

In this paper we further investigate nucleosynthesis in the outflows from
NS--NS/NS--BH merger remnant accretion discs around BHs, by applying a
nuclear reaction network on thermodynamic trajectories extracted from
fully time-dependent, long-term hydrodynamic simulations of disc
outflows. Our aim is to carry out a systematic study of the dependence
of the $r$-process production on system parameters such as
disc mass or viscosity, and on additional ingredients such
as nuclear physics inputs to the reaction network or the feedback from
nuclear heating on the disc dynamics. Our main conclusion is that
disc outflows from NS binary mergers can in principle produce both the
light and heavy $r$-process elements, without necessarily requiring
additional contributions from the dynamical ejecta.

The paper is organized as follows. Section \ref{sec-methods} describes
the hydrodynamic models, extraction of thermodynamic trajectories, and
the properties of the nuclear reaction network. Results and analysis
are presented in section \ref{sec-results}. Finally, section
\ref{sec-discussion} summarizes our findings and discusses broader
astrophysical implications.

\section{Computational Method}
\label{sec-methods}

\subsection{Disc evolution and thermodynamic trajectories}
\label{sec-trajectories}

We evolve NS merger remnant accretion discs around BHs 
using the approach described
in FM13, \citet{Metzger&Fernandez14}, and \citet{FKMQ14}. The equations of hydrodynamics
are solved numerically using FLASH3 \citep{fryxell00,dubey2009}. The public
version of the code has been modified to include
the equation of state of \citet{timmes2000} with abundances 
of nucleons and $\alpha$ particles ($^4$He) in nuclear statistical equilibrium (NSE), 
angular momentum transport due to an $\alpha$ viscosity \citep{shakura1973}, 
and the pseudo-Newtonian potential of \citet{artemova1996}. In addition,
neutrino emission and absorption due to charged-current weak interactions
on nucleons are included through energy- and lepton number
source terms using a leakage scheme for cooling and a disc-adapted
lightbulb approach for self-irradiation, with simple optical-depth 
corrections (see FM13 and \citealt{Metzger&Fernandez14} for details). 

\begin{figure*}
\centering
\includegraphics*[height=3in]{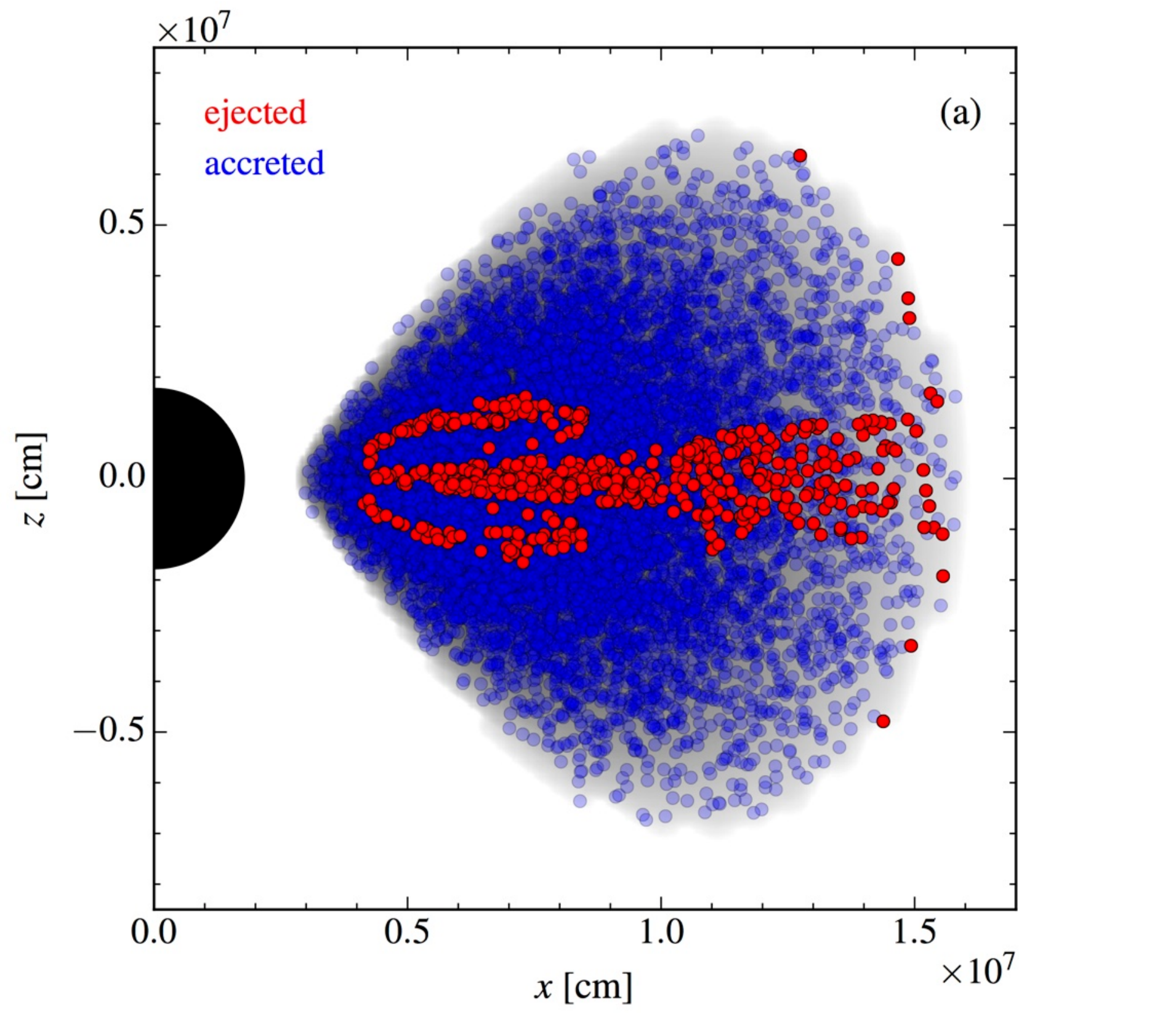}
\includegraphics*[height=3in]{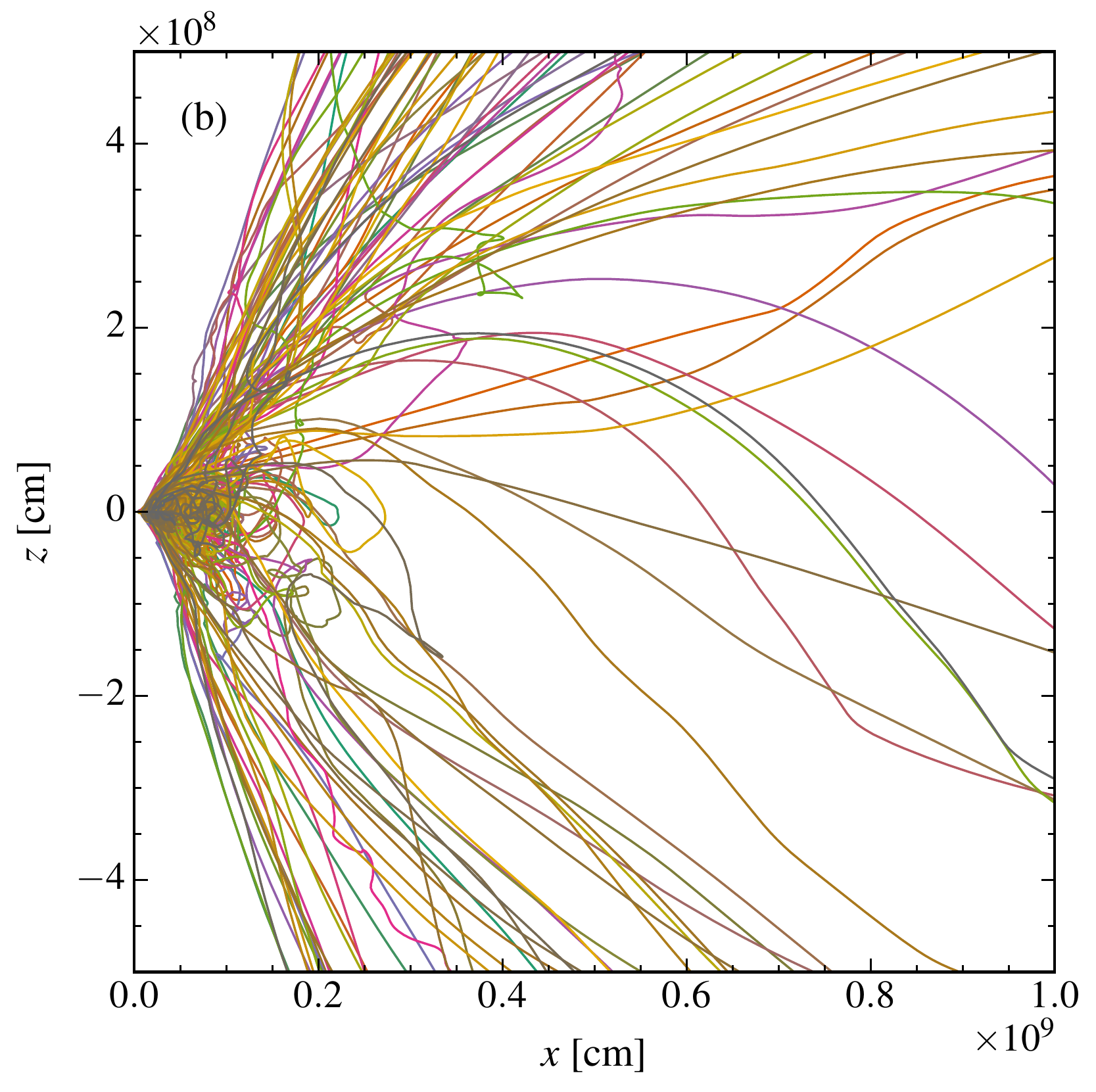}
\caption{Left: initial density field (grey) and particle
  distribution in model S-def. Particles that reach a distance of
  $2\times 10^9$~cm from the BH are considered to be ejected (red),
  with most of the remainder being accreted (blue). Right:
  paths followed by ejected particles, showing the convective
  character of the disc outflow (only 20\% of the outflow trajectories
  are shown, for clarity).}
\label{f:ic_paths}
\end{figure*}

Discs start from an equilibrium initial condition and are evolved in
axisymmetric spherical polar coordinates over thousands of orbits at
the initial density maximum (physical time of several seconds). The
majority of the mass loss is produced once the disc reaches the
advective state, in which viscous heating and nuclear recombination
forming $\alpha$ particles are unbalanced by neutrino cooling
\citep{Metzger+09a}. Heating due to neutrino absorption makes only a
minor contribution to the total mass ejection if a BH sits at the
centre (FM13, \citealt{just2014}). Total ejected masses can range from
a few percent to more than $\sim 20\%$ of the initial disc mass for a
highly-spinning BH.

Thermodynamic trajectories of the outflow are obtained by inserting
passive tracer particles in the hydrodynamic simulations.  For each
model, $10^4$ equal mass particles are initially placed in the disc
with random positions that follow the mass distribution
(Fig.~\ref{f:ic_paths}a). Each particle records thermodynamic
variables including density, temperature, electron fraction and
nucleon abundances, as well as neutrino and viscous source terms. Some
of these quantities are used as input for the $r$-process nuclear
reaction network (\S\ref{sec-network}), while others are retained to
check for consistency.

In order to assess the effects of additional nuclear heating 
after the $\alpha$-recombination on the
disc dynamics, we carry out a few simulations in which we
include this extra source term. This heating is parametrized
analytically as a function of temperature by analysing an 
ensemble of trajectories in our baseline 
model (\S\ref{sec-res_nucheat}).

Relative to the BH accretion disc models of FM13,
we have corrected an error in the treatment of the nuclear binding
energy of $\alpha$ particles (as described in \citealt{Metzger&Fernandez14}), which results
in slightly lower mass ejection than reported in FM13.
Since \citet{FKMQ14} we have also corrected an error in the computation of the
charged-current weak interaction rates, whereby the neutron-proton
mass difference was missing in the argument of the Fermi-Dirac
integrals of the neutrino rates but not the antineutrino
rates (e.g., \citealt{bruenn85}). After correcting this error,
we obtain a slightly less neutron rich outflow distribution,
with a tail towards higher $Y_e$.
Nonetheless the average properties of the outflow remain relatively 
unchanged.

\subsection{Models Evolved}
\label{sec-models}

\begin{table}
\centering
\caption{Hydrodynamic disc models evolved.\label{tab-models}
Columns from left to right are: initial torus mass, black 
hole mass, radius of initial torus density peak, initial
electron fraction, initial entropy, viscosity parameter, 
BH spin parameter, and amplitude of parametrized
nuclear heating (equation~\ref{eq-nucheat}).
}
\begin{tabular}{lcccccccc}
\hline\hline
{Model}&
{$M_{\rm t0}$} &
{$M_{\rm BH}$} &
{$R_0$} &
{$Y_{e0}$} &
{$s_0$} &
{$\alpha$} &
{$\chi$} &
{$\varepsilon$}\\
{} & \multicolumn{2}{c}{($M_\odot$)} & {(km)} & {} & {($k_{B}$/b)} & {} & {} & {}\\
\hline
S-def       & 0.03 & 3   &   50  & 0.10  & 8   & 0.03  & 0   & 0 \\ 
$\chi$0.8      &      &     &       &       &     &       & 0.8 &   \\
\noalign{\smallskip}
m0.01     & 0.01 & 3   &   50  & 0.10  & 8   & 0.03  & 0   & 0 \\
m0.10     & 0.10 &     &       &       &     &       &     &   \\
r75       & 0.03 & 3   &   75  &       &     &       &     &   \\
M10       &      & 10  &  150  &       &     &       &     &   \\
y0.05     &      & 3   &   50  & 0.05  &     &       &     &   \\
y0.15     &      &     &       & 0.15  &     &       &     &   \\
s6        &      &     &       & 0.10  & 6   &       &     &   \\
s10       &      &     &       &       & 10  &       &     &   \\
$\alpha$0.01     &      &     &       &       & 8   & 0.01  &     &   \\
$\alpha$0.10     &      &     &       &       &     & 0.10  &     &   \\
\noalign{\smallskip}
$\varepsilon$1.0      & 0.03 & 3   &   50  & 0.10  & 8   & 0.03  & 0   & 1.0\\
$\varepsilon$0.1      &      &     &       &       &     &       &     & 0.1\\
$\varepsilon$10.0     &      &     &       &       &     &       &     & 10.0\\
\hline\hline
\end{tabular}
\end{table}

Table~\ref{tab-models} presents all the models studied in this
paper. Following FM13, the baseline configuration (model S-def)
consists of a BH with mass $M_{\rm BH}=3M_\odot$, and an equilibrium
torus with disc mass $M_{\rm t0} = 0.03M_\odot$, radius at density
peak $R_0 = 50$~km, constant specific angular momentum, constant
initial entropy $s_0 = 8$~k$_{\rm B}$ per baryon, constant electron
fraction $Y_{e0} = 0.1$, viscosity parameter $\alpha = 0.03$ and BH
spin $\chi = 0$. We also include a model ($\chi 0.8$) with BH spin
parameter $\chi = 0.8$ to bracket the range of outcomes expected for an
NS--NS merger.

The remaining models vary one parameter at a time, mirroring the
simulations in FM13. In addition, we evolve three models that include
nuclear heating after $\alpha$ recombination in the hydrodynamics
(\S\ref{sec-trajectories}).  All models are evolved for 3000 orbits
measured at $R_0$, which amounts to several seconds of physical time,
in order to approach convergence in mass ejection
(the disc material left over interior to $r=2\times 10^9$~cm amounts to
a few percent of the initial disc mass).

\subsection{Nuclear Reaction Network}
\label{sec-network}

The initial disc temperature is of the order of a few MeV, so that
nearly all the ejecta consist initially of neutrons and protons.
Once the temperature drops to $\sim 10$~GK,
most of the free protons recombine with neutrons to form
$^4$He, as favoured by NSE 
under neutron-rich conditions ($Y_e\sim 0.2$).
The nuclear binding energy released by forming $^4$He plays
an important role in unbinding the material and is
included in the hydrodynamic simulations as described in
Sec.~\ref{sec-trajectories}.

In order to follow the change in abundances of nuclear species in the
disc outflow, we use a large nuclear reaction network for all tracer
particles that reach a distance of $2\times 10^9$~cm from the centre
of the BH.  The nuclear reaction network includes 7360 nuclei from
nucleons to $^{313}$Ds, and reaction rates including $\alpha$-decays,
$\beta$-decays, charged-particle reactions, neutron captures and their
inverse photo-dissociation interactions, as well as spontaneous,
$\beta$-delayed, and neutron-induced fission and the
corresponding fission yields~\citep[see][for a detailed
  description of the nuclear physics input
  used]{Mendoza-Temis.Wu.ea:2015}. In addition, we include
charged-current neutrino interactions on free nucleons as described in
Sec.~\ref{sec-trajectories}.

For each tracer particle, we start the network calculation from the
last moment when the temperature equals 10~GK.  For a small fraction
($\lesssim 10$\%) of the particles the temperature is always below
10~GK, because they are initially located at the outer edge of the
disc (e.g. Fig.~\ref{f:ic_paths}a).  For these particles, we instead
follow the nucleosynthesis evolution from the beginning of the disc
simulation. The initial composition of each trajectory is determined
by NSE given the temperature, density, and $Y_e$. The density
evolution is obtained at early times by interpolating the values
provided by the simulation. Once the network evolution reaches the end
of the disc simulation time $t_f$, we assume homologous expansion such
that $\rho(t)=\rho(t_f)\times (t_f/t)^3$ subsequently.  The
temperature evolution is interpolated from the simulation as long as
the temperature satisfies $T > T_s\equiv 6$~GK.  For $T<T_s$, the
temperature is evolved with the network following the same method used
in \cite{Freiburghaus+99} by including the energy sources from the
viscous and the neutrino heating for $t<t_f$ provided by the
hydrodynamic simulations, and the nuclear energy release due to
reactions calculated directly by the network. We have checked that
the final overall abundances are not sensitive to the choice of $T_s$.
This suggests that, differently to what happens in dynamical ejecta,
$r$-process heating in disc ejecta does not strongly influence the
abundances because the total energy released by this channel is much
smaller than the internal energy of the ejecta.  This is consistent
with the findings of \cite{just2014}. However, we discuss in
Sec.~\ref{sec-res_nucheat} the potential impact of nuclear heating on
the disc dynamics and on specific abundance features.

Since the $r$-process typically involves the nuclear physics
properties of very neutron-rich nuclei that are not experimentally
known, the outcome of nucleosynthesis calculations is subject to
theoretical uncertainties in the modelling of these properties. In the
case of dynamical ejecta, it has been shown that both the
neutron-capture and $\beta$-decay rates can substantially influence
the positions and heights of the $r$-process peaks (e.g.,
\citealt{Mendoza-Temis.Wu.ea:2015}, \citealt{Eichler:2014kma}). Our
standard set of calculations uses neutron-capture and
photo-dissociation rates based 
on nuclear masses from the Finite Range Droplet Model (FRDM)
\citep{Moller:1993ed} supplemented by the $\beta$-decay rates of
\cite{Moller:2003fn} for all disc models listed in
Table~\ref{tab-models}. To address the so far unexplored impact of
nuclear physics on disc ejecta, we have performed calculations using
neutron capture and photodissociation rates based on the Duflo--Zuker mass model
with 31 parameters \citep[DZ31;][]{Duflo:1995ep} and $\beta$-decay rates
from \cite{Marketin:2015gya}, see Sec.~\ref{sec-nuc_input}. 

\section{Results}
\label{sec-results}

Neutrino cooling in the disc is strong during the first
$\sim 10-100$~ms, largely balancing viscous heating. As the disc
spreads, neutrino cooling shuts off and energy deposition by viscous
heating and $\alpha$ recombination make the disc highly
convective. The resulting outflow peaks on a time-scale of
$\sim 1$~s, measured at a large enough radius that a significant fraction 
of the material will become gravitationally unbound from 
the system upon expansion ($r\sim 10^9$~cm).
Weak interactions drive $Y_e$ towards its equilibrium value, which in
the inner disc can be as high as $\sim 0.4$. However, most of this
material is accreted on to the BH. The bulk of the wind material arises
from regions of the disc midplane and on its back side relative to the
BH (Fig.~\ref{f:ic_paths}a), and are thus only moderately influenced
by neutrino interactions. A component close to the BH at high altitude
is mixed into the wind, increasing the average electron fraction,
particularly when the BH spin in high \citep{FKMQ14}.

During the initial expansion at high temperature
(10~GK$\gtrsim T\gtrsim 5$~GK), charged-particle reactions
dominate the nucleosynthetic flow to form seed nuclei.
At $T\approx 5$~GK, seed nuclei abundances peak at 
$A\approx 50$ ($N\approx 30$) and $A\approx 80$ 
($N\approx 50)$.
Heavier nuclei are only synthesized when $T\lesssim 5$~GK
by subsequent neutron capture processes and $\beta$-decays.
Given the importance of this temperature threshold, it is 
useful to examine different quantities when this point is reached.
For each trajectory, we denote the value of a given quantity $X$ at $T = 5$~GK by
\begin{equation}
\label{eq:X5_def}
X_5 \equiv X\bigg|_{T = 5\textrm{ GK}}.
\end{equation}
If $T=5$~GK is reached multiple times due to convective motions 
(Fig.~\ref{f:ic_paths}b), 
$X_5$ is chosen as the value at the last time
when $T=5$~GK.
It is also informative to measure quantities in the disc
outflow at large radius, where the effects of the BH
and convection are less likely to affect the properties
of the outflow. We thus denote the value of a given 
quantity $X$ at $r = 10^9$~cm by
\begin{equation}
\label{eq:X9_def}
X_9 \equiv X\bigg|_{r = 10^9\textrm{ cm}}.
\end{equation}
Finally, we denote the average of a quantity over 
all trajectories by
\begin{equation}
\label{eq:X5ave_def}
\bar X_s \equiv \frac{\sum_i X_{s,i}}{N},
\end{equation}
where $N$ is the total number of trajectories, $i$ labels each
trajectory, and the subscript $s$ can be $5$ or $9$. Note that since
all trajectories have the same mass, this is a mass average.

\subsection{Ejecta properties and nucleosynthesis in the baseline model}
\label{sec-res_s_def}

\begin{figure}
\includegraphics[width=\columnwidth]{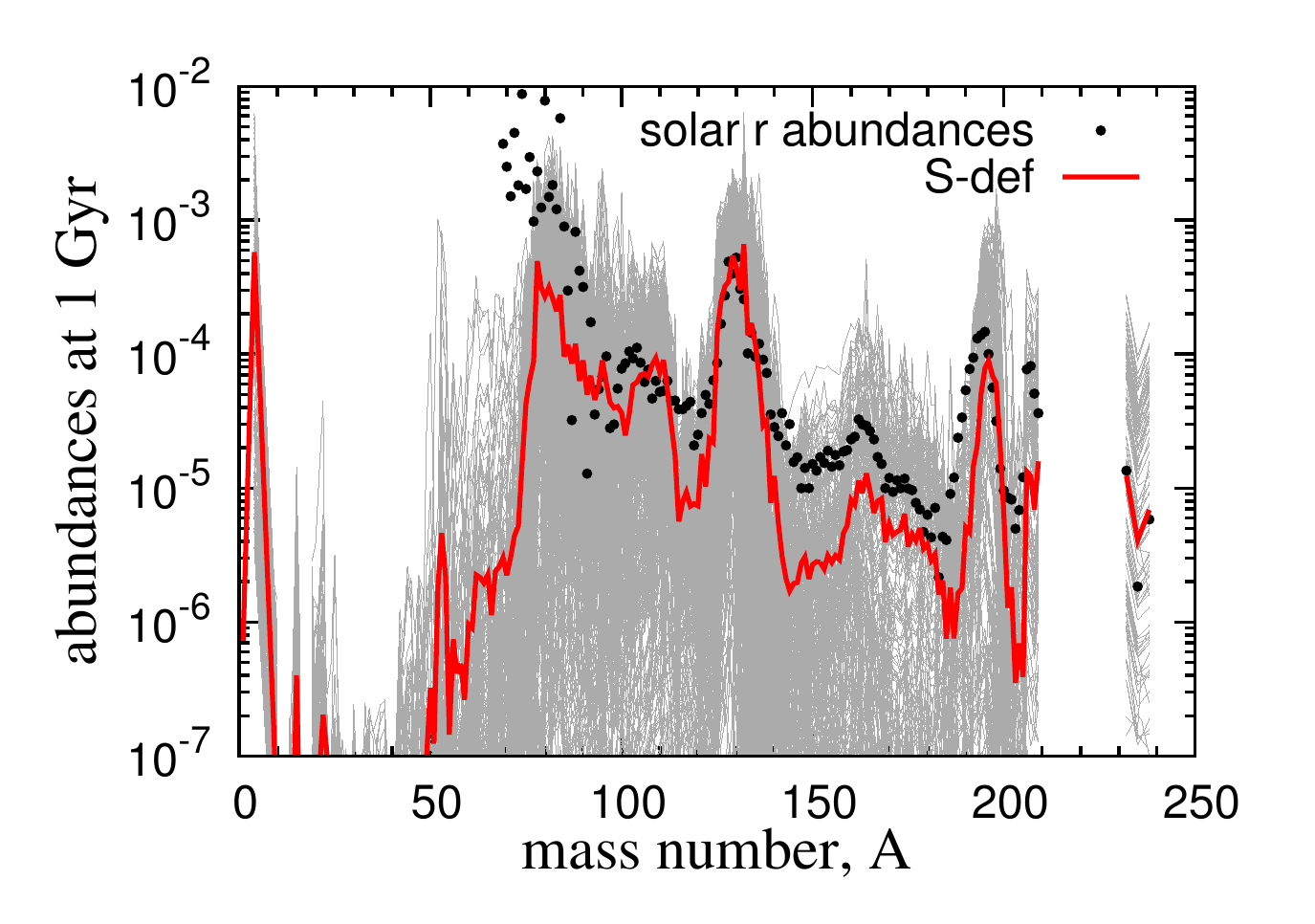}
\caption{Abundances at $1$~Gyr as a function of mass number $A$ for
  the baseline model S-def. Shown are the average abundances over all
  trajectories (thick red curve) and results for individual particles
  (thin grey curves).  Black dots show the Solar system
  $r$-abundances~\citep{Cowan:1998nv} scaled to match the averaged
  abundances around the second peak.\label{fig-intd_sdef}}
\end{figure}

\begin{figure}
\includegraphics[width=\columnwidth]{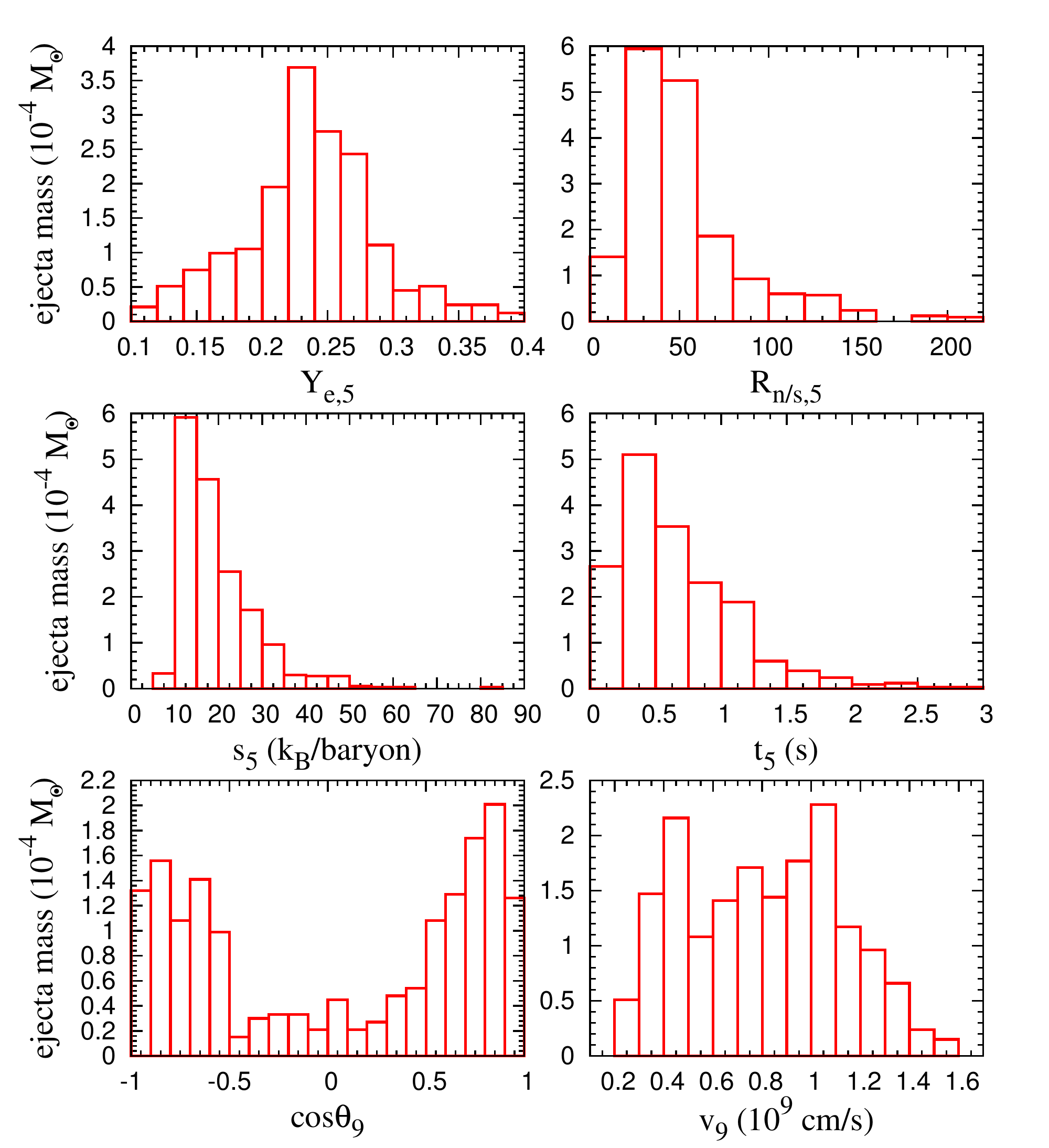}
\caption{Mass distribution of the disc outflow from the 
baseline model S-def, as a function of electron fraction $Y_{e,5}$, 
neutron-to-seed ratio $R_{n/s,5}$, entropy $s_5$, 
time $t_5$, polar angle $\theta_9$, and radial velocity $v_9$. Subscripts have
the meanings defined in equations~(\ref{eq:X5_def}) and (\ref{eq:X9_def}).} 
\label{fig-histo}
\end{figure}

The $r$-process abundance distribution obtained with the baseline
model S-def is shown in Fig.~\ref{fig-intd_sdef}.  A breakdown of
the thermodynamic- and kinematic properties of the ejecta as a
function of various quantities is illustrated in
Fig.~\ref{fig-histo}. The particle-averaged properties are also
summarized in Table~\ref{tab:models-nuc}.  The disc outflow has a
broad distribution of electron fraction spanning the range
$0.1\leq Y_{e,5}\leq 0.4$, with a mean at $\bar Y_{e,5} \simeq 0.24$.
The entropy peaks at $s_5=10$~k$_{\rm B}$ per baryon, with a long
tail extending out to $\sim $30~$k_B$ per baryon, and the
neutron-to-seed ratio $R_{n/s,5}$ lies in the range 20--100. In terms
of kinematics, most of the outflow reaches $5$~GK within 2~s and
its angular distribution peaks at $\sim\pm 45^\circ$--$60^\circ$
from the disc
equatorial plane. Ejecta velocities lie in the range
$(0.2\textrm{--}1.6)\times 10^9$~cm~s$^{-1}$, much slower than the typical
values for the early phase of the dynamical ejecta (see,
e.g.,~\citealt{just2014}).  With such a broad distribution of
$Y_{e,5}$, the resulting average nucleosynthetic abundances are in
overall agreement with the Solar system $r$-process distribution from
$A\sim 80$ to U and Th, as shown in Fig.~\ref{fig-intd_sdef}, despite
a large spread among individual trajectories.

\begin{figure*}
\center
\includegraphics[width=\textwidth]{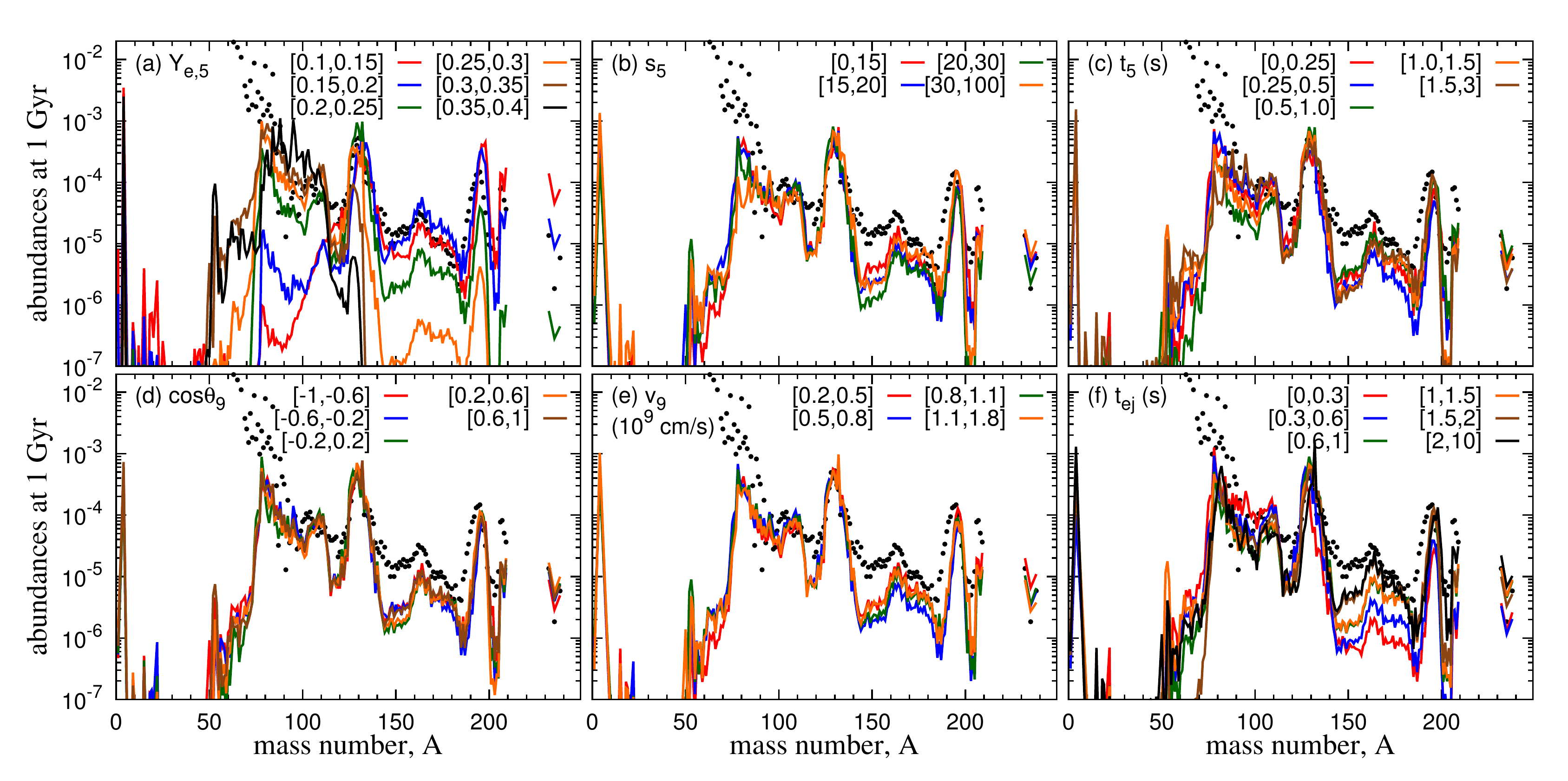}
\caption{Dependence of the abundances on different ejecta properties,
  for the baseline model S-def. Each panel shows average abundances as
  a function of mass number, binned into intervals of the variable
  shown in the upper-left corner: electron fraction (a), entropy (b),
  time at $T=5$~GK (c), polar angle (d), radial velocity (e), and
  ejection time (f), with subscripts defined in
  equations~(\ref{eq:X5_def}) and (\ref{eq:X9_def}). The interval
  values are shown in the corresponding legends.}
\label{fig-intd_sep_sdef}
\end{figure*}

In order to clarify the origin of the spread in abundances for
individual particles, we bin trajectories into intervals of electron
fraction $Y_{e,5}$, entropy $s_5$, time $t_5$, polar angle $\theta_9$,
and radial velocity $v_9$, with the results shown in
Fig.~\ref{fig-intd_sep_sdef}.  The nucleosynthesis results are most
sensitive to the electron fraction (see
Fig.~\ref{fig-intd_sep_sdef}a).  Most of the abundances above the
rare-Earth peak ($A\gtrsim 150$) come from ejecta with
$Y_{e,5}\lesssim 0.2$.  Trajectories with $Y_{e,5}$ in the range
0.1--0.3 contribute to the production of the second peak
($A\simeq 130$). As for the abundances around $A\lesssim 110$, they
are dominantly from the ejecta with $Y_{e,5}\gtrsim 0.25$. Fission
plays a role only for ejecta with $Y_e \lesssim 0.15$ and consequently
represents a minor contribution to the final abundances.

Interestingly, the nucleosynthesis outcome is not sensitive to
the entropy intervals in the ejecta as shown in Fig.~\ref{fig-intd_sep_sdef}b. 
This is mainly
because the different entropy bins have nearly the same 
electron fraction distribution, with only a minor trend of
lower entropy having lower peaked $Y_{e,5}$ distributions.
As a result, we do not recover the entropy dependence obtained in 
parametrized nucleosynthesis studies such as \cite{Lippuner:2015gwa}. 
The abundance patterns are rather insensitive to 
other quantities such as the 
polar angle $\theta_9$ of each particle (see Fig.~\ref{fig-intd_sep_sdef}c-e).

\begin{figure}
\includegraphics[width=\columnwidth]{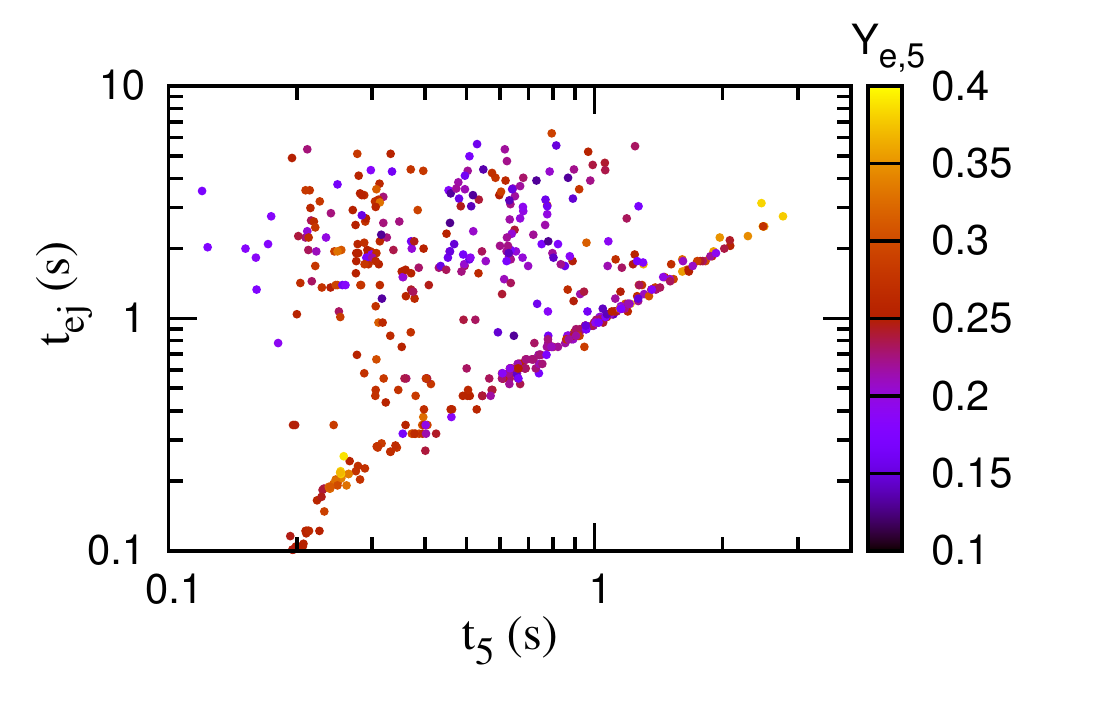}
\caption{Ejection time $t_{\rm ej}$ as a function of time $t_5$
  (equation~\ref{eq:X5_def}), for all trajectories from the baseline
  model S-def. Each point is colored by the electron fraction at time
  $t_5$.}
\label{fig-tej_t5}
\end{figure}

The disc outflow is strongly convective out to late times
(Fig.~\ref{f:ic_paths}b), unlike the ejecta from the neutrino-driven
wind in core-collapse supernovae or the dynamical ejecta of
NS--NS/NS--BH mergers, for which the density and temperature of the
ejecta monotonically decrease as the medium expands. Particles in the
disc outflow may in fact go through several convective cycles before
being ejected at temperatures of $\sim 1$~GK.  To quantify the effect
of this process on the nucleosynthesis, we define the ejection time
$t_{\text{ej}}$ of each particle as the time in the simulation after
which the radial position only increases monotonically with time.
This quantity is shown in Fig.~\ref{fig-tej_t5} as a function of
$t_5$ for all particles in the S-def model, with each point colored by
the electron fraction $Y_{e,5}$.

Fig.~\ref{fig-tej_t5} clarifies several aspects of the ejecta
behaviour.  First, there is a substantial amount of material having
much larger $t_{\rm ej}$ than $t_5$ for $t_5\lesssim 1$~s.  Before
ejection, these particles are subject to strong convective motions in
low-temperature regions ($T<5$~GK) at a late phase in the disc
evolution.  Secondly, the electron fraction $Y_{e,5}$ of trajectories
ejected at the earlier times is higher.  This occurs because neutrino
irradiation plays a larger role in unbinding this early ejecta and
thus naturally raises its $Y_{e,5}$ to higher values.  Consequently,
the rare-earth and third peak abundances are much lower for matter
being ejected at earlier time than for later ejecta, as shown in
Fig.~\ref{fig-intd_sep_sdef}f.  In particular, the abundance of
Lanthanides can differ by almost a factor of 10 compared to the
latest particles.  This is consistent with the results of
\citet{FKMQ14}, who on the basis of the $Y_e$ distribution of the
ejecta suggested that the wind contains a Lanthanide-free ``skin'' --
even in the case of a non-spinning BH -- which can generate a small
amount of blue optical emission if not obstructed by neutron-rich
dynamical ejecta \citep{KFM15}.

Note also that a very strong $A=132$ abundance spike around the second
peak is formed for ejecta with $2.0<t_{\rm ej}<10$~s.  This feature
and its relation to convective motions is the subject of
\S\ref{sec-res_anomaly}.

\subsection{Dependence on model parameters}
\label{sec-res_model}

\begin{table*}
\centering
\begin{minipage}{13cm}
  \caption{Average ejecta properties for models listed in
    Table~\ref{tab-models}\label{tab:models-nuc}.  Columns from left
    to right show model name, electron fraction, entropy, time,
    expansion times-cale $\tau_{\rm exp} = {\bar r}/{\bar v}_r$,
    radius, velocity, final average mass number, percentage of ejected
    mass, percentage of ejected tracer particles that
    contain both $Y_{A=132}>10^{-3}$ and
    $Y_{A=128(136)}<0.1Y_{A=132}$, and the total mass fraction of
    lanthanides and actinides ($57\leq Z\leq 71$ and
    $89\leq Z\leq 103$), $X_{\rm Lan+Act}$, at
    $t=1$~d.
    Subscripts and averages are defined in
    equations~(\ref{eq:X5_def})-(\ref{eq:X5ave_def}).  Only
    trajectories that reach $r=2\times 10^9$~cm at the end of the
    simulation are considered.  }
\begin{tabular}{lcccccccccc}
\hline
Model  &$\bar Y_{e,5}$&$\bar s_5$      &$\bar t_5$&$\bar\tau_{\rm exp,5}$&$\bar r_5$     &
$\bar v_{r,9}$ &$\langle\bar A\rangle_f$&$M_{\rm ej}/M_{t0}$&132-peak
& $X_{\rm Lan+Act}$\\
       &         &($k_{B}$/b)&(s)  &(ms)        &($10^7$cm)&$10^9$cm~s$^{-1}$&                    &\%                 &\%  & \% \\
\hline
S-def & 0.237 & 19.9 & 0.65 & 81  & 3.84 & 0.81 & 119 & 5.67  & 11.8 & 4.30\\
$\chi$0.8  & 0.282 & 19.9 & 0.81 & 82  & 4.56 & 1.04 & 103 &18.05  & 13.2 & 2.01\\
\noalign{\smallskip}
m0.01 & 0.202 & 20.4 & 0.43 & 89  & 3.18 & 0.91 & 137 & 5.32  & 16.5 & 8.32\\
m0.10 & 0.261 & 19.1 & 1.04 & 104 & 4.90 & 0.93 & 109 & 5.00  & 10.6 & 3.02\\
M10   & 0.228 & 20.2 & 0.53 & 227 & 6.48 & 0.62 & 119 & 6.29  & 24.2 & 5.49\\
r75   & 0.231 & 17.6 & 0.77 & 117 & 4.94 & 0.73 & 122 &14.80  & 22.3 & 5.40\\
s10   & 0.260 & 18.8 & 0.60 & 91  & 4.29 & 0.75 & 107 & 7.19  & 16.4 & 2.34\\
s6    & 0.203 & 20.8 & 0.73 & 96  & 3.75 & 0.86 & 136 & 4.22  & 13.0 & 0.91\\
$\alpha$0.01 & 0.321 & 26.1 & 2.66 & 100 & 2.88 & 0.70 & 91  & 1.36  & 0    & 0.25\\
$\alpha$0.10 & 0.206 & 18.1 & 0.16 & 52  & 4.56 & 1.23 & 130 &13.53  & 2.2  & 8.85\\
y0.05 & 0.230 & 20.5 & 0.72 & 110 & 3.93 & 0.91 & 122 & 5.00  & 13.6 & 5.12\\
y0.15 & 0.252 & 19.4 & 0.66 & 102 & 4.44 & 0.87 & 112 & 5.72  & 19.2 & 3.26\\
\noalign{\smallskip}
$\varepsilon$1.0  & 0.230 & 14.9 & 0.41 & 111 & 4.28 & 0.85 & 118 & 11.39 & 5.1  & 4.40\\
$\varepsilon$0.1  & 0.228 & 17.5 & 0.58 & 95  & 3.90 & 0.77 & 122 & 6.08  & 15.3 & 4.77\\
$\varepsilon$10.0 & 0.216 & 14.5 & 0.28 & 131 & 4.24 & 2.38 & 123 & 16.11 & 0    & 5.09\\
\hline
\end{tabular}
\end{minipage}
\end{table*}

\begin{figure*}
\center
\includegraphics[width=\textwidth]{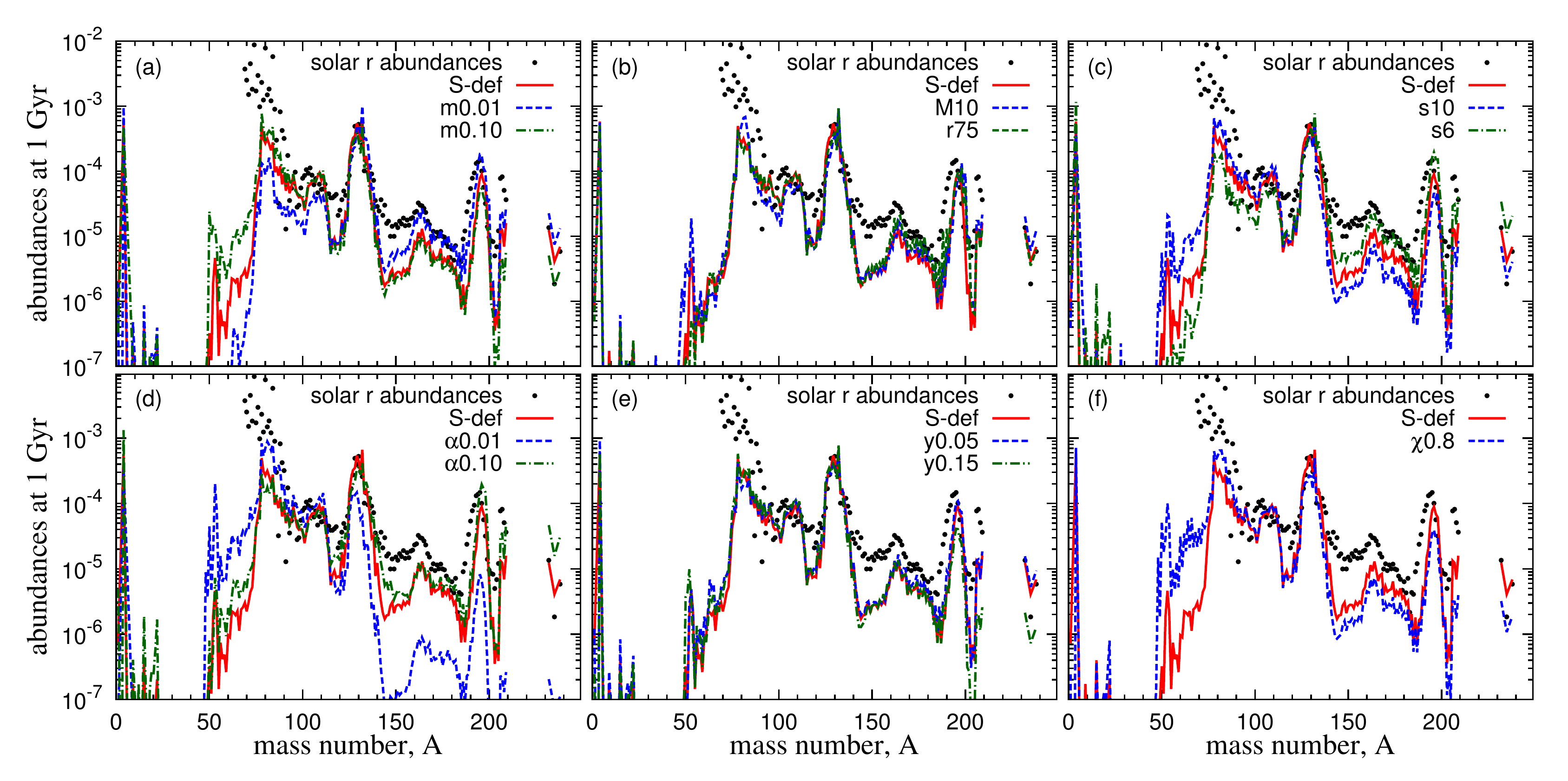}
\caption{Dependence of average disc outflow abundances on 
disc mass (a), BH mass and disc radius (b), initial entropy (c),
viscosity parameter (d), initial electron fraction (e), and BH spin (f).
Model parameters are shown in Table~\ref{tab-models},
with nucleosynthesis results summarized in Table~\ref{tab:models-nuc}.
Note that the Solar system $r$-abundances are scaled to
match the second peak abundances of S-def model only. When
comparing to abundances of other models, they should be further re-scaled.} 
\label{fig-intd_model_comp}
\end{figure*}

The effect of changing various disc parameters on the average
nucleosynthetic abundances of the outflow is shown in 
Fig.~\ref{fig-intd_model_comp} and summarized in 
Table~\ref{tab:models-nuc}.
We note that the Solar system $r$-abundances shown in
Fig.~\ref{fig-intd_model_comp} are scaled to
match the second peak abundances of S-def model only. When
comparing to abundances of other models, they should be further re-scaled.

The largest impact of changing disc parameters
is obtained when varying the $\alpha$ viscosity 
parameter (see Fig.~\ref{fig-intd_model_comp}d).
Since the disc evolution times-cale is governed by the
viscous time $t_{\rm visc}\propto \alpha^{-1}$,
models with lower (higher) viscosity have a longer 
(shorter) $\bar t_5$ relative to the S-def model.
The corresponding $\bar Y_{e,5}$ is raised to higher (lower)
values by weak interactions. As a result, model $\alpha$0.10
($\bar Y_{e,5}=0.206$) has about a factor of 2
higher abundances of third peak and heavier elements compared 
to S-def ($\bar Y_{e,5}=0.237$). At the other end, 
model $\alpha$0.01 ($\bar Y_{e,5}=0.321$) displays a 
relatively large first peak $(A\approx 80)$
while the abundances at and above the second peak are
reduced by roughly an order of magnitude.

The initial disc mass and entropy have a moderate impact
(see Fig.~\ref{fig-intd_model_comp}a and c).
For discs with initially larger (smaller) entropy,
the temperature is higher (lower). As a result,
higher (lower) positron capture rates give rise 
to a higher (lower) value of $\bar Y_{e,5}$.
Similarly, for a disc with initially larger (smaller)
mass, $\bar Y_{e,5}$ is also higher (lower) because
the disc temperature is higher (lower) for fixed
entropy, disc radius, and BH mass due to the higher (lower) density.
Smaller changes are obtained when varying the initial 
electron fraction (see Fig.~\ref{fig-intd_model_comp}e). 
Higher (lower) initial $Y_e$ leads to a slightly larger (smaller) 
third peak and trans-lead abundances.
This insensitivity originates in the fact that, 
while the outflow itself does not achieve $\beta$-equilibrium 
(c.f. Fig.~6 of \citealt{Fernandez&Metzger13}), weak
interactions nonetheless modify $Y_e$ significantly
from its initial value.

Regarding BH spin, a higher value of this parameter
results in a smaller innermost stable
circular orbit, leading to more gravitational energy release
in the form of neutrinos and viscous heating (e.g., \citealt{FKMQ14}).
As a result, $\bar Y_{e,5}$ is higher
in model $\chi 0.8$ than in the baseline S-def model,
slightly hindering the nucleosynthesis of heavy nuclei
for both the second and third peaks while
enhancing the first peak abundances
as shown in Fig.~\ref{fig-intd_model_comp}f.

Changing the mass of the BH and radius of the disc do
not lead to significant differences given our parameter choices. 
Fig.~\ref{fig-intd_model_comp}b
shows that there are minor differences between models
M10 and S-def. The similarities between them can be
attributed to the very similar compactness, $M_{\rm BH}/R_0 = 0.07$ and
$0.06 M_\odot$~km$^{-1}$ for M10 and S-def, respectively, which should correlate
with the disc temperatures (and thus strength of weak interactions) if the 
medium is in near hydrostatic equilibrium. In contrast, model r75 has a 
somewhat lower compactness $0.04M_\odot$~km$^{-1}$ and
shows a slightly larger abundance of elements with $A > 130$.

The dependence of the ejecta mass distribution on flow
quantities -- such as polar angle and velocity -- for
models with different parameters is generally similar to
that for the baseline model S-def 
shown in Fig.~\ref{fig-histo}.

We conclude that some nucleosynthesis features -- particularly abundances
of elements with $A\lesssim 130$ -- are quite robust against
variations in astrophysical disc parameters.
On the other hand, the abundances of rare-earth peak, third peak and
heavier elements are more sensitive to those parameters that alter
either the disc evolution time-scale or the temperature.

\subsection{Anomalously high abundance of $A=132$}
\label{sec-res_anomaly}

Fig.~\ref{fig-intd_model_comp} shows that all models we have 
explored, except $\alpha$0.01 and $\alpha$0.10, display a
peculiarity in the abundances at the second peak: the abundance of 
$A=132$ elements ($Y_{A=132}$)
exceeds that of the second Solar system $r$-process peak 
at $A = 128-130$ ($Y_{A=128-130}$)
We find that this spike in $Y_{A=132}$ arises primarily from
late-time ejecta (see, e.g., Fig.~\ref{fig-intd_sep_sdef}f). 
Table~\ref{tab:models-nuc} shows that material satisfying
$Y_{A=132}>10^{-3}$ and $Y_{A=128(136)}<0.1 Y_{A=132}$
can contribute more than 10\% to the total ejected mass.
This would cause a discrepancy with
the observed solar abundance ratio
$Y_{A=132}/Y_{A=130}\approx 0.5$ if we
were to assume that the disc ejecta dominates the 
galactic $r$-process production of second peak elements.

The origin of this anomaly can be
traced back to late-time, low-temperature convection in the disc
outflow.  For ejecta satisfying $t_{\rm ej}\gg t_5$, neutron-capture
processes operate once the temperature decreases to $\lesssim 2$--3~GK. Once
neutrons are exhausted, the produced nuclei at $N=82$, mainly isotones
with $A=128$--130, start to decay to the stability
valley. If convection brings matter back to regions of
$T\gtrsim 2$--3~GK on time-scales shorter than the half-lives of long
lived Sn isotopes: $^{128}$Sn
($t_{1/2} = 59.07$~min), $^{129}$Sn ($t_{1/2} = 2.23$~min), and $^{130}$Sn
($t_{1/2} = 3.42$~min), additional neutron captures can occur due to
the release of neutrons by photodissociation. This results in a pile up
of material in the double-magic nucleus $^{132}$Sn that itself has
a $\beta$-decay half-life of $t_{1/2} = 39.7$~s. 

The model with the smallest viscosity parameter ($\alpha$0.01) does
not display this anomaly because on one hand, the strength of
convection is strongly reduced relative to the baseline value.
Correspondingly, most of the ejecta from this model satisfy
$t_{\rm ej}\sim t_5$. On the other hand, most of the ejecta in this
model contains $Y_{e,5}\gtrsim 0.25$ so that little nuclei with
$A>130$ can be produced.  At the opposite end, the model with the
highest viscosity ($\alpha$0.10) also displays a much smaller
abundance anomaly than S-def.  In this case, if the viscous time is
shorter than the $r$-process time scale $\lesssim 1$~s, each
convection episode results in an incomplete $r$-process that does not
use all available neutrons. Only the very last ejection episode
results in a complete $r$-process. 
Therefore, only little trajectories contain
the abundance anomaly.

The above discussion illustrates the close interplay between
convection and $r$-process nucleosynthesis. 
Convection operating on
time-scales longer than the $r$-process time-scale
of $\lesssim 1$~s will reheat the $r$-process
products during their decay back to stability
and produce anomalies in the abundance
distribution. 

\subsection{Effect of nuclear heating}
\label{sec-res_nucheat}

The hydrodynamic disc models discussed so far do not self-consistently
include the nuclear energy released by charged-particle reactions
involving nuclei heavier than $^4$He, in addition to the energy
released by the $r$-process. These contributions are only included in
post-processing with the nuclear network and hence can only affect the
temperature.  Charged-particle reactions can release a nuclear energy
$\sim 3$~MeV/nucleon through processes like
$20$n+15$\alpha\rightarrow^{80}$Zn. This can give rise to a heating
rate $\sim 10^{19}$~erg~s$^{-1}$, assuming that such reactions occur
on a time-scale of $\sim 0.1$~s. It can become comparable to the
viscous heating rate in the same temperature regime, and has the
potential to change not only the nucleosynthesis but also the disc
dynamics.  The energy released during the $r$-process is smaller in
comparison, amounting to $\sim 10^{18}$~erg~s$^{-1}$, but can
potentially also play a role at late times.  Given the abundance
anomaly discussed in the previous section, and its sensitivity to
convection in the disc, we explore here the extent to which
additional sources of heating can introduce changes in the dynamics
and nucleosynthesis results.

\begin{figure}
\includegraphics[width=\columnwidth]{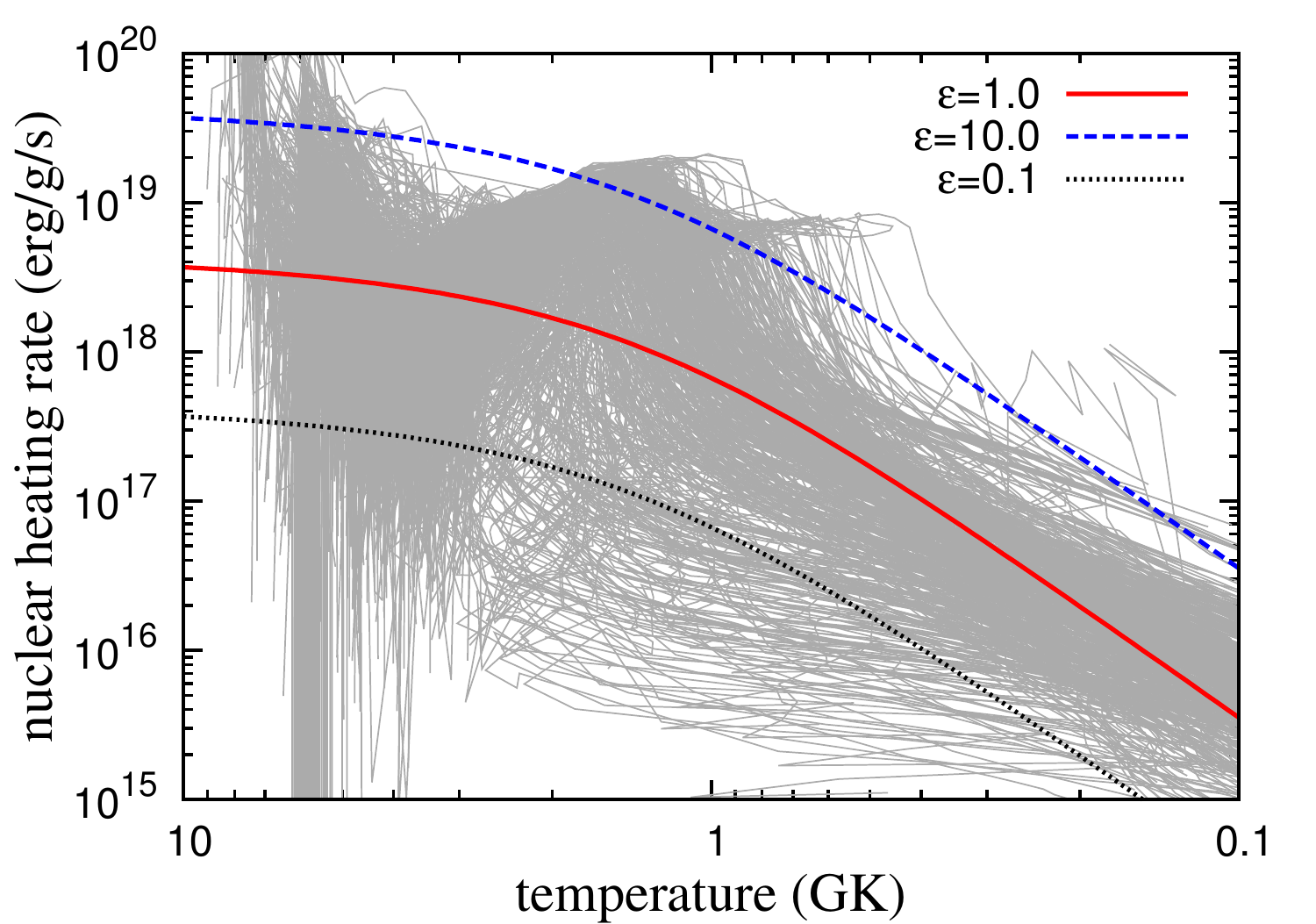}
\caption{Nuclear heating rate as a function of temperature for all
  trajectories from the baseline model S-def (grey curves).  Also
  plotted is an analytic fit to the average heating rate
  (equation~\ref{eq-nucheat}) for three choices of the heating amplitude
  $\varepsilon$, as shown in the legend.  }
\label{fig-QT}
\end{figure}

In order to include these additional heating channels into
hydrodynamic models, we have parametrized it out of trajectories from
the S-def model using a method similar to that described
in~\citet{just2014}.  From our post-processed nucleosynthesis
calculation, we estimate the particle-averaged nuclear heating rate
$\langle\dot q\rangle$ as a function of the average temperature
$\langle T\rangle$ by calculating
\begin{equation}
\label{eq:qave_def}
\langle\dot q(\tilde t)\rangle \equiv \frac{\sum_i\dot q_i(\tilde t)}{N}
\end{equation}
and 
\begin{equation}
\label{eq:Tave_def}
\langle T(\tilde t)\rangle\equiv \frac{\sum_i T_i(\tilde t)}{N},
\end{equation}
where $\dot q_i(\tilde t)=\dot q_i(t-t_{0,i})$ and
$T_i(\tilde t)=T_i(t-t_{0,i})$ are the nuclear energy generation rate
and temperature of each trajectory $i$, respectively, and $t_{0,i}$ is
the time at which the network integration begins for each trajectory
(\S\ref{sec-network}).  Following \cite{Korobkin+12}, we find that the
function $\langle{\dot q}\rangle(\langle T\rangle)$ can be
approximated by
\begin{equation}
\label{eq-nucheat}
\langle{\dot q}\rangle(\langle T\rangle)
= 2.5\times 10^{19}\,\varepsilon\left[
\frac{1}{\pi}\arctan\left(\frac{\langle T\rangle}{1.1\textrm{ GK}}\right)
\right ]^{5/2}\,\textrm{ erg g}^{-1}\textrm{ s}^{-1},
\end{equation}
with $\varepsilon\simeq 1.0$.
Fig.~\ref{fig-QT} compares the result from  
equation~(\ref{eq-nucheat}) with the heating rates from individual
particles in the S-def model.

We evolve three additional hydrodynamic models that include
this additional heating rate using a range of values for the
heating amplitude $\varepsilon=\{0.1,1,10\}$ to compensate for the crudeness
of the approximation (models $\varepsilon$0.1, $\varepsilon$1.0 and
$\varepsilon$10.0 in 
Table~\ref{tab-models}). The heating term
is added to the energy equation once $T \leq 6$~GK if
$v_r >0$. The latter condition is used to prevent heating in
flows that move back towards the BH.

Table~\ref{tab:models-nuc} shows that the additional heating can
enhance the ejected mass up to a factor of 2 when using
$\varepsilon\gtrsim 1.0$. 
The angular distribution can also be affected. In the model $\varepsilon$10.0, the mass ejection becomes more isotropic
rather than peaks at high latitudes.

\begin{figure}
\includegraphics[width=\columnwidth]{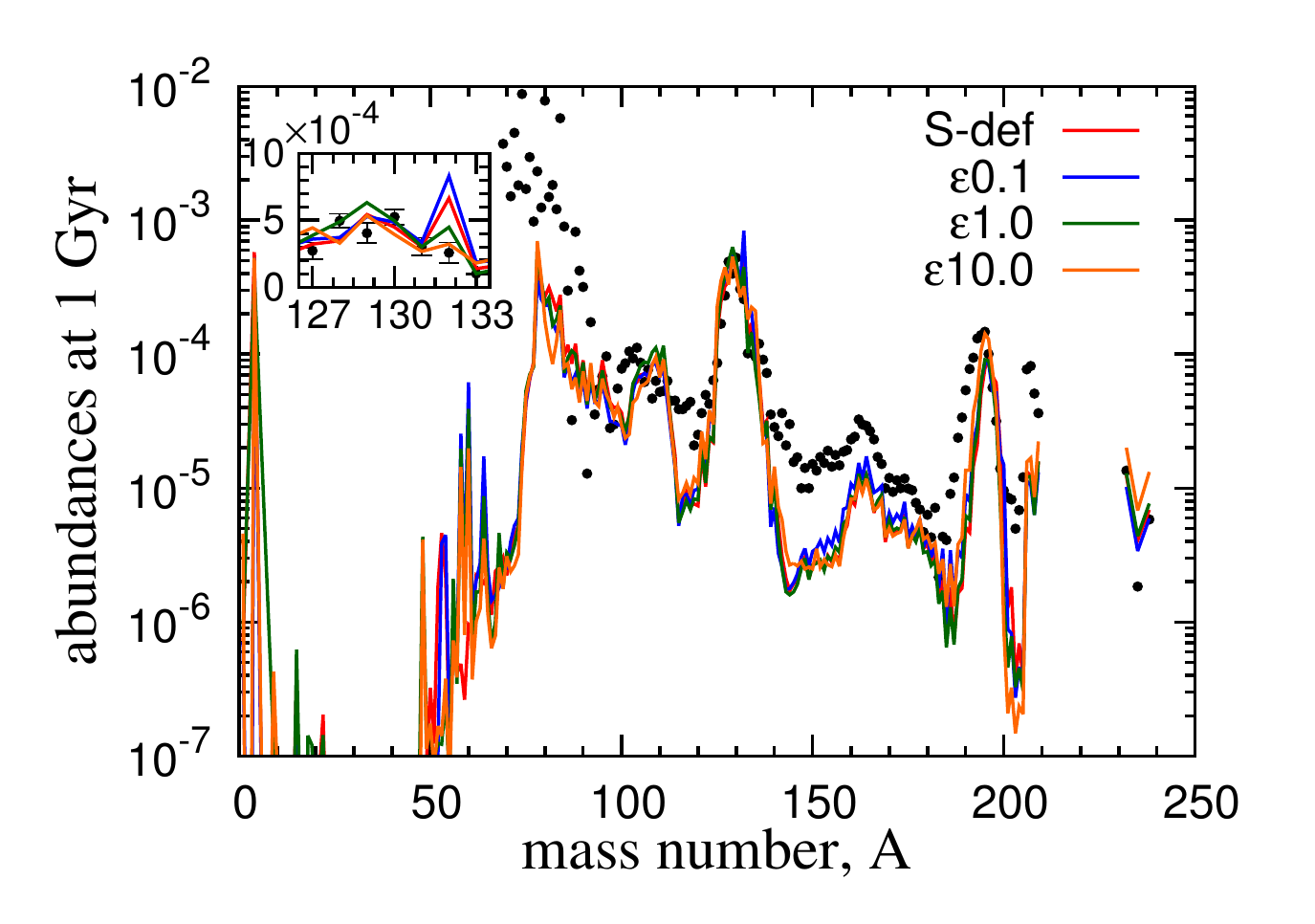}
\caption{Dependence of the average disc outflow abundances on the
  amplitude $\varepsilon$ of the parametrized radioactive heating
  rate (equation~\ref{eq-nucheat}), relative to the baseline model S-def
  (c.f. Table~\ref{tab-models}). The inset shows the amplitude of the
  anomalous abundance peak at $A=132$ in comparison with the
    solar $r$-abundances including observational uncertainties.}
\label{fig-intd_s_h}
\end{figure}

Nucleosynthesis results are shown in Fig.~\ref{fig-intd_s_h}.  The
abundances of third peak and heavier elements are slightly increased
relative to the baseline model S-def.  Interestingly, the $A=132$
abundance anomaly is reduced in model $\varepsilon$1.0 and completely
vanishes in model $\varepsilon$10.0, as shown in the inset of
Fig.~\ref{fig-intd_s_h}.  For the $\varepsilon$1.0 model, this
suppression occurs because the disc evolution becomes faster as a
result of the extra heating, so that the amount of late-time ejecta
affected by convection is reduced.  For model $\varepsilon$10.0, the
nuclear heating is so strong that convection is strongly suppressed.

Our results indicate that including all nuclear heating sources may be
necessary to fully understand the ejecta dynamics and the details of
the nucleosynthesis outcome. Alternatively, the dynamics of the disc
when transporting angular momentum via magnetohydrodynamic stresses
might be different enough to erase the $A=132$ abundance anomaly
without the need for enhanced nuclear heating.  The properties of
convective motion in a purely hydrodynamical $\alpha$-disc may differ
substantially from those of MHD turbulence (e.g.,
\citealt{Balbus&Hawley02}).

\subsection{Impact of nuclear physics input}
\label{sec-nuc_input}

\begin{figure}
  \includegraphics[width=\columnwidth]{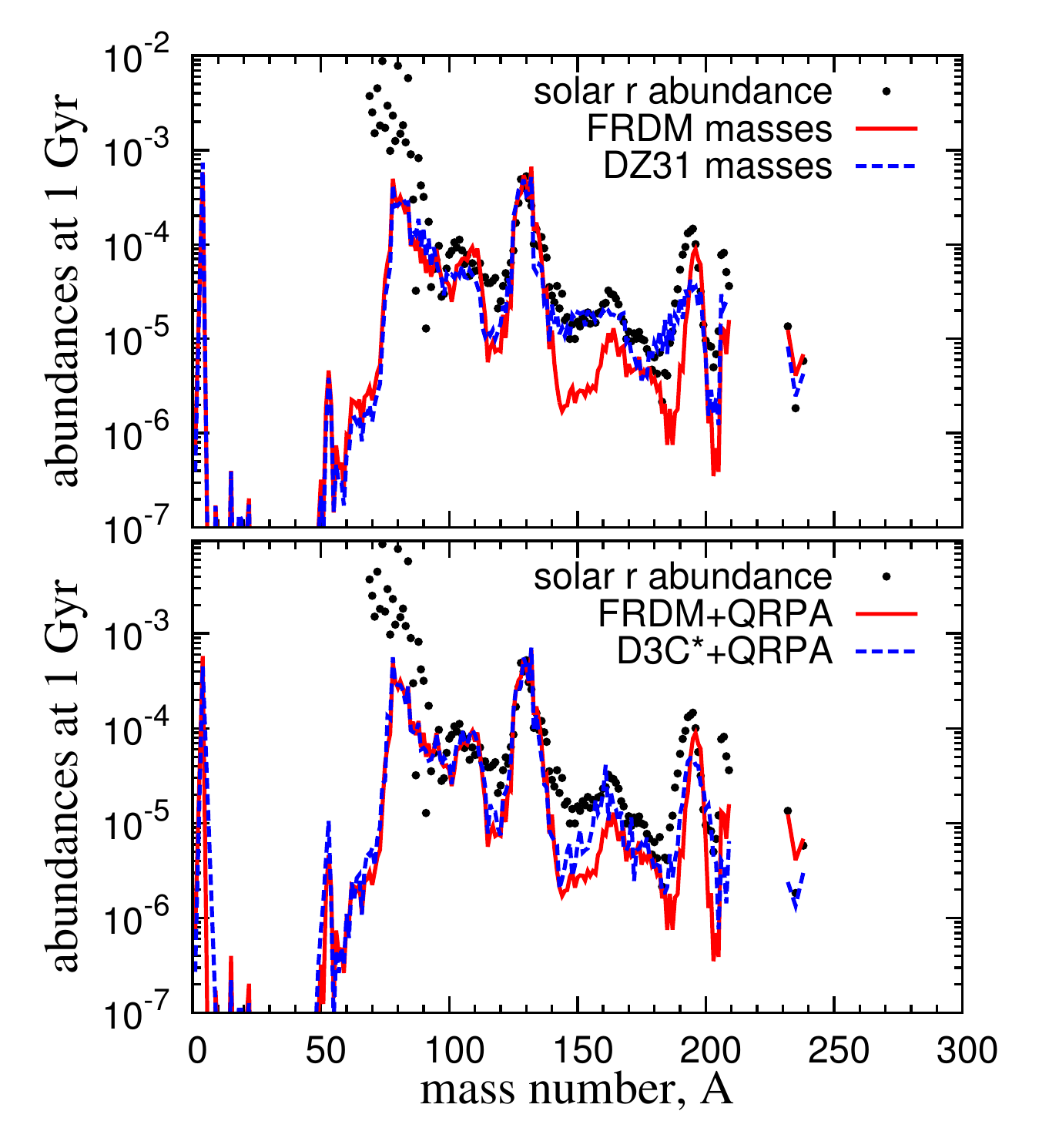}
  \caption{Average abundances in the outflow from the baseline model
    S-def for different nuclear mass models (top) and $\beta$-decay
    rates (bottom).  See \S\ref{sec-nuc_input} for details.}
  \label{fig-nucinput}
\end{figure}

In order to explore the effect of varying the nuclear physics inputs
on the outflow nucleosynthesis, we have performed two additional
network calculations on the baseline model S-def using different
neutron-capture and photo-dissociation rates, as well as $\beta$-decay
rates.  Alternative neutron capture and photodissociation rates are
computed with the nuclear mass model of Duflo-Zuker with 31 parameters
(DZ31; \citealt{Duflo:1995ep}), as in
\cite{Mendoza-Temis.Wu.ea:2015}. To explore the impact of
$\beta$-decay rates, we alternatively employ those
of~\cite{Marketin:2015gya} and \cite{Moller:2003fn}.

The results of these additional calculations are shown in
Fig.~\ref{fig-nucinput}.  The first and second $r$-process peaks are
robust against changes in nuclear physics inputs, as they are produced
primarily by trajectories with high $Y_{e,5}$, whose nucleosynthesis
paths are close to the stability valley.  For these trajectories, the
relative change of neutron separation energy is small between the
different mass models, and the predictions from two sets of
$\beta$-decay rates are similar.  In contrast, nuclear physics inputs
can strongly affect the abundances of rare-earth and third peak
elements -- produced by low $Y_{e,5}$ ejecta -- to a similar or even
larger degree than the change due to astrophysical parameters
discussed in \S\ref{sec-res_model}. In particular, the
troughs after the $A\sim130$ peak and before the 
$A\sim 195$ peak may be due to deficiencies in the FRDM
mass model~\citep{Winteler.Kaeppeli.ea:2012}. A more
detailed analysis of the effect of nuclear physics inputs
will be reported separately.

\subsection{Comparison with abundances of metal-poor stars}
\label{sec-mpstars}

We have thus far restricted the comparison of disc nucleosynthesis
abundances to the Solar system $r$-process distribution.  Observations
of $r$-process abundances in metal poor stars
($[\text{Fe/H}] = \log(N_{\rm Fe}/N_{\rm H})_*- \log(N_{\rm Fe}/N_{\rm
  H})_\odot\lesssim -2.5$)
suggest that the $r$-process operated at early times in the Galaxy
\citep{Sneden+08}. While a number of metal-poor stars show
robust relative abundances in the region around the
rare-Earth peak ($56\lesssim Z\lesssim 72$) that are similar to the
solar $r$-abundance distribution, larger relative variations exist
among lighter elements ($30\lesssim Z\lesssim 50$).  Moreover, the
ratio between light-to-heavy elemental abundances can vary from star
to star by more than a factor of 10. This has lead to
suggestions of the existence of more than one astrophysical site producing
the $r$-process elements in the early Galaxy
\citep{Qian&Wasserburg07,Sneden+08}.

\begin{figure}
\includegraphics[width=\columnwidth]{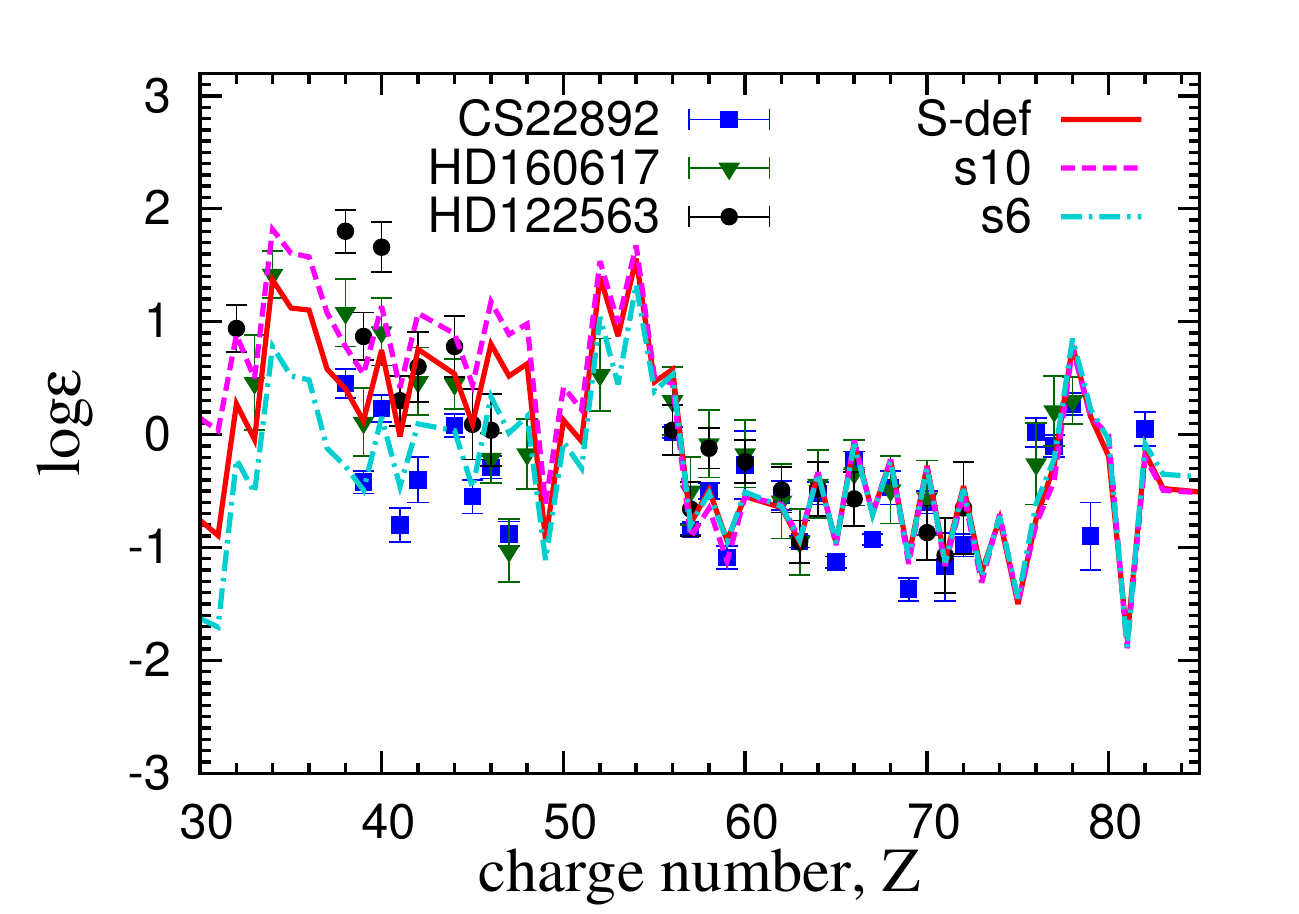}
\caption{Comparison of average elemental abundances in the
outflows from the baseline model S-def and models with
different initial entropy (s6 and s10) to observed
abundances in three metal-poor stars:
CS22892-052 \citep{Sneden:2003zq}, 
HD160617 \citep{Roederer:2012ve}, 
and HD122563 \citep{Roederer:2012dr}. 
Abundances 
are re-scaled to the Eu abundance of 
CS22892-052. Here $\log\varepsilon(Z)=\log(N_Z/N_{\rm 1})+12$,
where $N_Z$ is the abundance of an element with charge number
$Z$.
}
\label{fig-mpstars}
\end{figure}

Fig.~\ref{fig-mpstars} shows the comparison of elemental abundances
from three models (S-def, s6, s10) to the observed abundances in three
metal-poor stars: CS22892-052 \citep{Sneden:2003zq}, HD160617
\citep{Roederer:2012ve}, and HD122563 (\citealt{Honda:2006kp},
\citealt{Roederer:2012dr}), normalized to the Eu abundance of
CS22892-052. These three stars show large differences in the ratio of
light-to-heavy elemental abundances. If we consider disc outflows
alone, changing the initial entropy of the disc can lead to
qualitative agreement with the observations in these three metal-poor
stars across the whole range of neutron capture elements.  As
shown in Sec.~\ref{sec-res_model}, changing other disc parameters such
as the initial disc mass or the adopted $\alpha$-viscosity can also
result in similar or larger changes in abundances. This suggests that
if the initial conditions in NS--NS/NS--BH mergers lead to variations in
the remnant configurations, the resulting disc outflows may account
for different observed ratios of light and heavy elemental $r$-process
abundances in metal poor stars. Therefore, the observed
variations may not only be due to different combinations of
dynamical and disc ejecta (as suggested by \citealt{just2014}), but
may arise due to variations of the merger remnant disc properties.

\section{Summary and Discussion}\label{sec-discussion}

We have studied the production of $r$-process elements in the
disc outflows from remnant accretion tori around BHs formed
in NS-NS/NS-BH mergers. We used tracer particles in long-term,
time dependent hydrodynamic simulations of these discs 
to record thermodynamic and kinematic quantities. The resulting
trajectories were then post-processed with a dynamic $r$-process
nuclear reaction network. Our results can be summarized as follows:
\newline

\noindent 1. -- Outflows from merger remnant discs around BHs can
	            robustly generate light $r$-process elements with 
                $A \lesssim 130$, regardless of the astrophysical
                parameters of the disc or the nuclear physics inputs 
                employed (Fig.~\ref{fig-intd_sep_sdef},
                Table~\ref{tab:models-nuc}). 
\newline

\noindent 2. -- The yield of elements with $A > 130$ is most sensitive
                to the type and magnitude of the angular momentum
                transport process (Fig.~\ref{fig-intd_model_comp}d) and 
                on the nuclear physics inputs employed
                in the nuclear reaction network (Fig.~\ref{fig-nucinput}). 
                Given the physics 
                employed in our hydrodynamic models, the $\alpha$ viscosity
                parameter is a key factor determining the abundance of
                third peak and heavier elements. Other parameters such as
                the disc mass or initial entropy have a relatively
                smaller impact 
                on the abundances.
\newline
                
\noindent 3. -- We have identified a spike in the abundance of
	  	        $A=132$ elements that arises whenever the disc outflow
                is highly convective, as is the case when using reasonable
                choices for the disc parameters
                (Fig.~\ref{fig-intd_sep_sdef}f).  
                This feature can be erased
                if the disc evolution is fast or if the heating rate in the
                disc is very low, so that convection is suppressed.
\newline
                
\noindent 4. -- Inclusion of energy deposition from charged-particle reactions
                beyond $^4$He recombination can affect the ejecta dynamics and
                nucleosynthesis to the point where the
                $A=132$ abundance anomaly disappears (Fig.~\ref{fig-intd_s_h}). 
                Alternatively, the 
                processes responsible for controlling angular momentum transport and
                the thermodynamics of the disc (e.g. MHD and neutrino transport) can
                have a sensitive nucleosynthetic impact.  The properties of
                convective motions in an hydrodynamical $\alpha-$disc may differ
                substantially from those of MHD turbulence
                (\citealt{Balbus&Hawley02}).
\newline

\noindent 5. -- The comparison with abundances
                  observed in metal-poor stars shows that if the disc
                  outflows contribute dominantly to the NS-NS/BS-BH
                  ejecta, different initial configurations of the disc
                  may account for the variation of light-to-heavy
                  abundance ratio seen in these stars.
\newline               

Our results together with those of~\citet{just2014} show that disc
outflows are fundamental to understand $r$-process nucleosynthesis in
mergers. For
cases in which little dynamical ejecta is generated,
{\it the disc outflow alone can contribute substantially to the
heavy $r$-process element enrichment}, even while
producing proportionally more elements with $A < 130$.
This result reinforces the general view that 
NS-NS/NS-BH mergers are the primary astrophysical
site for heavy $r$-process elements.  It is also important
because the presence of high-$Y_e$ dynamical ejecta 
in NS-NS mergers
is uncertain theoretically, due in part to the dependence of shock-heated
polar ejecta on the NS radius and equation of state.  

The early ejection of material with high $Y_e$
in the disc outflow has the potential to generate
a blue peak in the kilonova, since it will
generally reside on the outer layers of the ejecta \citep{FKMQ14,KFM15}.
This nonetheless depends on the viewing directions
close to the rotation axis to be relatively free
of Lanthanide-rich dynamical ejecta material, which
is usually the case for BH-NS mergers \citep{FQSKR15,KFM15}. 

The net kilonova contribution of systems studied
is Lanthanide-rich, as inferred from Table~\ref{tab:models-nuc}.
The work of \citet{kasen2013} shows that even a
mass fraction of $\sim 10^{-2}$ in Lanthanides
can increase the optical opacity by at least an
order of magnitude relative to that of iron-like
elements. Nearly all of our models have a Lanthanide
mass fraction bigger than $1\%$ and thus while an outer
Lanthanide-free ``skin" is usually obtained, the bulk of the wind will
lead to an infrared transient.  More promising in this respect is the
possible onset of a neutron-powered precursor if a small fraction of
material escapes quickly enough to freeze-out before neutrons are
captured by heavy seeds \citep{metzger2015_n-precursor}. Simulations
at much higher resolution than currently available are needed to
resolve this question.  A long-lived NS remnant can also increase the
quantity of high-$Y_e$ ejecta, producing a more prominent blue
component of the kilonova (\citealt{Metzger&Fernandez14}).

A crucial improvement to our calculation would involve
obtaining trajectories from an accretion disc outflow in
which angular momentum transport -- and the associated
energy dissipation -- is carried out by MHD stresses.
Such a calculation can differ from ours in a number of
ways. First, the change in the entropy due to viscous heating
in an $\alpha$-viscosity model is likely different in 
an MHD disc, with the associated change in the equilibrium 
$Y_e$ that the weak interactions try to achieve. Secondly,
the amount of mass ejected and the associated $Y_e$ distribution
can change, altering the relative amounts of heavy- and light
$r$-process elements in the outflow composition. Finally,
the kinematic properties of the wind can change, in particular
the velocity, which controls the expansion time, and the
angular distribution, which is associated with the level
of neutrino irradiation of the ejecta (material on the
equatorial plane is more effectively shadowed from neutrinos
than material leaving at high latitudes).

Our calculations would also benefit from better neutrino 
transport, although the magnitude of the difference introduced
can be comparable to that due to the spin of the BH (e.g., 
compare the results of \citealt{FKMQ14} and \citealt{just2014}).
A more thorough, self-consistent treatment of nuclear heating would also make
our calculations more realistic. In this respect, it is
worth noting that the results of \citet{just2014} do not appear 
to show the abundance spike at $A=132$ that we obtain in many of our
models. 
We surmise that this difference arises due to their
inclusion of a single species of heavy nucleus ($^{54}$Mn) in
the equation of state, which partially
accounts for the energy production beyond 
$\alpha$ formation.

Finally, it is important to consider the combined evolution
of disc and dynamical ejecta in computing the net $r$-process 
yield from NS-NS/NS-BH mergers. Part of the dynamical ejecta
is gravitationally bound, and mixes with the accretion disc,
increasing the neutron-richness of the disc and therefore
lowering the peak of the $Y_e$ distribution of the disc
outflow \citep{FQSKR15}.

\section*{Acknowledgements}

We gratefully thank Lutz Huther for his help
during the initial stage of this work. 
We also thank Almudena Arcones, 
Kenta Hotokezaka for helpful discussions,
Friedel Thielemann, Andreas Bauswein, Thomas Janka,
Oliver Just, Luciano Rezzolla, Luke Roberts, and Ian Roederer
for their valuable comments.
MRW and GMP acknowledge support from the Deutsche
Forschungsgemeinschaft through contract SFB~1245, the Helmholtz
Association through the Nuclear Astrophysics Virtual Institute
(VH-VI-417), and the BMBF-Verbundforschungsprojekt number 05P15RDFN1.
RF acknowledges support from the University of California Office of
the President, and from NSF grant AST-1206097.  BDM gratefully
acknowledges support from NASA grants NNX15AU77G (Fermi), NNX15AR47G
(Swift), and NNX16AB30G (ATP), NSF grant AST-1410950, and the Alfred
P. Sloan Foundation.  The software used in this work was in part
developed by the DOE NNSA-ASC OASCR Flash Center at the University of
Chicago.  This research used resources of the National Energy Research
Scientific Computing Center (repository m2058), which is supported by
the Office of Science of the U.S.  Department of Energy under Contract
No. DE-AC02-05CH11231. Some computations were performed at
\emph{Carver} and \emph{Edison}.  This research also used the
\emph{Savio} computational cluster resource provided by the Berkeley
Research Computing program at the University of California, Berkeley
(supported by the UC Berkeley Chancellor, Vice Chancellor of Research,
and Office of the CIO).

\bibliographystyle{mn2e}
\bibliography{old_refs,rodrigo}

\label{lastpage}
\end{document}